\documentclass[12pt, apj]{emulateapj}
\usepackage{epsfig}
\usepackage{graphicx}
\usepackage{wrapfig}
\newcommand{\beq}	{\begin{equation}}
\newcommand{\eeq}	{\end{equation}}
\newcommand{\beqa}{\begin{eqnarray}}
\newcommand{\eeqa}{\end{eqnarray}}
\newcommand{\beqs}	{\begin{displaymath}}
\newcommand{\eeqs}	{\end{displaymath}}
\newcommand{\beqas}	{\begin{eqnarray*}}
\newcommand{\eeqas}	{\end{eqnarray*}}

\def\bit{\begin{itemize}}
\def\eit{\end{itemize}}

\newcommand{\eee}	{$^{-3}$}
\def\simlt{\lower.5ex\hbox{$\; \buildrel < \over \sim \;$}}
\def\simgt{\lower.5ex\hbox{$\; \buildrel > \over \sim \;$}}

\newcommand\eff{{\rm eff}}
\newcommand\leff{\lambda_\eff}
\newcommand\lj{\lambda_{\rm J}}
\newcommand\mH{m_{\rm H}}
\newcommand\ssb{\sigma_{\rm SB}}

\begin{document}
\author{
Charles E. Hansen \altaffilmark{1}, Richard I. Klein \altaffilmark{1,2}, 
Christopher F. McKee \altaffilmark{1,3}, and Robert T. Fisher \altaffilmark{4}
}
\submitted{The Astrophysical Journal, Accepted}

\altaffiltext{1}{Astronomy Department, University of California, Berkeley, CA 94720}
\altaffiltext{2}{Lawrence Livermore National Laboratory, Livermore, CA 94550}
\altaffiltext{3}{Physics Department, University of California, Berkeley, CA 94720}
\altaffiltext{4}{Physics Department, University of Massachusetts, Dartmouth, MA}

\title{Feedback Effects on Low-Mass Star Formation}

\begin{abstract}
Protostellar feedback, both radiation and bipolar outflows, dramatically affects the fragmentation and mass accretion from star-forming cores.  We use ORION, an adaptive mesh refinement (AMR) gravito-radiation-hydrodynamics code, to simulate low-mass star formation in a turbulent molecular cloud in the presence of protostellar feedback.  We present results of the first simulations of a star-forming cluster that include both radiative transfer and protostellar outflows.  We run four simulations to isolate the individual effects of radiation feedback and outflow feedback as well as the combination of the two.  We find that outflows reduce protostellar masses and accretion rates each by a factor of three and therefore reduce protostellar luminosities by an order of magnitude. This means that, while radiation feedback suppresses fragmentation, outflows render protostellar radiation largely irrelevant for low-mass star formation above a mass scale of 0.05 $M_{\odot}$.  We find initial fragmentation of our cloud at half the global Jeans length, around 0.1 pc.  With insufficient protostellar radiation to stop it, these 0.1 pc cores fragment repeatedly, forming typically 10 stars each.  The accretion rate in these stars scales with mass as predicted from core accretion models that include both thermal and turbulent motions; the accretion rate does not appear to be consistent with either competitive accretion or accretion from an isothermal sphere.  We find that protostellar outflows do not significantly affect the overall cloud dynamics, in the absence of magnetic fields, due to their small opening angles and poor coupling to the dense gas.  The outflows reduce the mass from the cores by 2/3, giving a core to star efficiency, $\epsilon_{\rm{core}} \simeq 1/3$.  The simulations are also able to reproduce many observation of local star-forming regions.  Our simulation with radiation and outflows reproduces the observed protostellar luminosity function.  All of the simulations can reproduce observed core mass functions, though we find they are sensitive to telescope resolution.  We also reproduce the two-point correlation function of these observed cores.  Lastly, we reproduce IMF itself, including the low-mass end, when outflows are included.

\end{abstract}

\section{Introduction}

The origin of the stellar initial mass function (IMF) is one of the most fundamental problems of star formation. The IMF can be described by single power law for stars above 0.5 $M_{\odot}$ \citep{Salpeter1955}, and a broken power law \citep{Kroupa2002} for stars below this mass.  Alternatively, it can be described as a log-normal distribution with characteristic mass $m_c = 0.2 M_{\odot}$ that joins with the Salpeter power law for stars above 1.0 $M_{\odot}$ \citep{Chabrier2005}.  Any theory of the IMF must explain both the functional form and the characteristic mass.  A tantalizing observational clue to the functional form lies in dust observations in star-forming regions \citep{Motte1998, Testi1998, Johnstone2000, Johnstone2001, Motte2001, Beuther2004, Stanke2006, Alves2007, Enoch2008, Sadavoy2010}.  These dust maps find many high density concentrations that are consistent with prestellar and protostellar cores.  When the mass of these cores is calculated, the core mass function (CMF) has the same functional form as the IMF, but with a higher characteristic mass, ranging from 0.2 $M_{\odot}$ to 1 $M_{\odot}$.  If each core is converted to a small number of stars with some efficiency, $0.2 < \epsilon_{\rm{core}} < 1.0$, the IMF can be recreated.  The actual conversion from observed core masses to stellar masses may not be so simple due to cores blending in projection \citep{Kainulainen2009a, Michel2011}, small cores that disperse before making stars \citep{PhilMyers2009, Padoan2011} and cores accreting mass over time \citep{Padoan2011}.  Nevertheless, the CMF likely plays a strong role in creating the IMF

The observed CMF provides support to core accretion theories of star formation \citep{Shu1977, McKee2003}, which start with a bound core and produce a single stellar system.  Simulations of turbulence find the functional form of the core mass function (log-normal plus power law) is the expected outcome of supersonic turbulence \citep{Padoan2002, Padoan2007}.  Analytic predictions of a turbulent density field with self-gravity can also reproduce this functional form \citep{Hennebelle2008, Hennebelle2009}.  The characteristic core mass is then the Jeans mass at some critical density and temperature.  However, choosing the correct density and temperature is problematic.  In purely isothermal turbulence, there is no characteristic Jeans mass.  As objects collapse and the density increases, the Jeans mass decreases.  There is no transition where this decrease in Jeans mass will stop without additional physics.  This means the core masses are either infinitely small or functions of the global Jeans mass of the host molecular cloud.  Observations are consistent with a universal IMF, however, even over a range of cloud Jeans masses \citep{Kroupa2002, Chabrier2003, Bastian2010}.  This means the characteristic core mass must be set by local physics, which isothermal turbulence cannot provide.  Star-forming regions are approximately isothermal because the thermal time scales are much shorter than the dynamical time scales, but there are ways to break this isothermality.

One approach is to focus on the coupling between gas and dust in star-forming environments \citep{Larson2005, Elmegreen2008}.  At low densities, gas-dust coupling is poor and the gas is theoretically slightly sub-isothermal (temperature decreases with increasing density).  At higher densities, gas and dust are well coupled and the gas is theoretically slightly super-isothermal.  This yields a critical density and temperature at the transition that can be converted into a Jeans mass.  This critical density, $\rho \sim 10^{-19}~\rm{g}~\rm{cm}^{-3}$,  is lower than the densities of large star-forming regions like Orion, however, and unlikely to explain the characteristic core mass in these regions.

One critical mass is the point when dust becomes opaque to its own thermal radiation \citep{Low1976}.  At that density, the gas will heat up and raise the Jeans mass, creating a minimum Jeans mass of fragmentation.  A barotropic simplification of this effect sets the mass in many simulations (e.g. \citealt{Bate2005, Bonnell2006, Offner2008, Bate2009b, Hennebelle2011}).  The density of this transition is extremely high ($\sim 10^{-13}~\rm{g/cm}^{3}$) \citep{Masunaga1998} and the resulting Jeans mass ($\sim 0.004 M_{\odot}$) is much lower than the characteristic core mass  \citep{Low1976, Whitworth2007}.  In addition, the barotropic approximation is inaccurate when compared to simulations that include dust radiation \citep{Boss2000, Krumholz2007a, Offner2009, Bate2009, Price2009, Tomida2010}.  The importance of dust radiation can be seen in \citet{Bate2009} and \citet{Price2009}, who found that the inclusion of radiation significantly suppresses the formation of brown dwarfs despite the near absence of protostellar luminosity in the simulations.

The most powerful break from isothermality comes from protostellar radiation.  Massive protostars are capable of heating an entire cloud \citep{Krumholz2007a, Cunningham2011, Myers2011, Krumholz2011}.  Low-mass stars do not have the same long range influence, but simulations show they can still dramatically reduce fragmentation in the disk and recover a 1 $M_{\odot}$ characteristic core mass \citep{Offner2009, Krumholz2011}.  Protostellar radiation does not create a unique critical density, but it does weaken the density dependence of the effective Jeans mass \citep{Bate2009}.

Given a core mass function, there is still the question of CMF to IMF efficiency, $\epsilon_{\rm{core}}$.  The primary mechanism for reducing the core mass is protostellar outflows. Stars of all masses show bipolar outflows during their formation \citep{Richer2000, Shepherd2003} and are recreated in MHD simulations with sufficient resolution \citep{Ciardi2010, Tomida2010}.  These outflows remove mass that would otherwise accrete onto stars, thereby reducing the final mass \citep{Matzner2000, Arce2006, Wang2010}.  Analytical estimates of mass loss from winds can fully explain the range of mass loss expected from observations $0.2 < \epsilon_{\rm{core}} < 1.0$ depending on the details of the cores and the outflows \citep{Matzner2000}.  The outflows travel beyond their stars of origin and deposit energy into parsec-scale turbulent motions.  Evidence suggests that molecular cloud turbulence appears on the scale of the entire cloud \citep{Ossenkopf2002, Brunt2009}, so is most likely driven by sources other than protostellar outflows.  Nonetheless, the dynamics on parsec scales can be significantly altered by outflows \citep{Norman1980, McKee1989, Li2006, Banerjee2007, Nakamura2007, Swift2008, Carroll2010, Arce2010, Wang2010}.

In this paper, we investigate the fragmentation of a parsec-scale molecular cloud into cores and then into stars. This requires refinement to capture the fragmentation and radiative transfer to fragment at the correct mass scale, similar to \citet{Offner2009}.  This also requires simulation of protostellar outflows, to capture the CMF to IMF efficiency, similar to the work of \citet{Cunningham2011} for high-mass star formation.  The goal of this paper is to explain both the observed CMF and the IMF while self-consistently finding $\epsilon_{\rm{core}}$.  In order to do this, we have performed the first simulation of a star-forming cluster to include both radiative transfer and protostellar outflows.

We describe our numerical method and simulation setup in \S 2.  In \S 3, we report the results of our simulations.  We discuss the implications of our results on star formation theory and compare to observations in \S 4.  We summarize our conclusions in \S 5.

\section{Simulations}

We perform four primary simulations with nearly identical initial conditions but different physics as controlled numerical experiments in order to isolate the importance of feedback effects.  These simulations all include hydrodynamics, gravity and basic sink particle physics \citep{Krumholz2004}, but may also include radiation \citep{Krumholz2007a} and/or sink particle outflows \citep{Cunningham2011}.  The simulations are shown in Table \ref{tab:simulations}.  Barotropic simulations are labeled with a B, simulations with radiation are labeled with R, and simulation with protostellar winds are labeled with W.  Simulation labels can contain multiple letters.

\begin{table}
\caption{Table of simulations}
\begin{tabular}{|c|c|c|c|}

  \hline
  Name & Thermal Physics & Winds? \\
  \hline
  B & Barotropic & No \\
  BW & Barotropic & Yes \\
  R & Radiation & No \\
  RW & Radiation & Yes \\

  \hline

\end{tabular}

\label{tab:simulations}
\end{table}

\subsection{Initial Conditions}

All simulations have the same initial conditions, also used in \citet{Offner2009}.  The initial gas temperature is $T_g = 10~K$, the box length is $L = 0.65$ pc and the average density is $\bar{\rho} = 4.46 \times 10^{-20}~\rm{g~cm}^{-3}$, corresponding to $n_H = 1.91 \times 10^{4}~\rm{cm}^{-3}$.  The total box mass is 185 $M_{\odot}$.  For radiative simulations, the radiation temperature, $T_r$ is initialized to 10 K.  The radiation energy density is thus $E = a T_r^4 = 7.56 \times 10^{-11}~\rm{erg}~\rm{cm}^{-3}$.

To obtain the turbulent initial conditions, we begin our simulations without self-gravity and apply velocity perturbations to an initially constant density field using the method described in \citet{MacLow1999}.  These perturbations correspond to a Gaussian random field with a flat power spectrum in the range $1 \leq k \leq 2$.  The application of these perturbations continues for three cloud crossing times and then stops.  At this point the turbulence follows a Burgers power spectrum, $P(k) \propto k^{-2}$, characteristic of supersonic hydrodynamic turbulence. The 3D turbulent Mach number is $\mathcal{M} = 6.6$, which gives a 3D rms velocity dispersion, $\sigma_v = 1.2~\rm{km/s}$.  With this Mach number the cloud is approximately virialized:
\begin{equation}
\alpha_{\rm{vir}} = \frac{5 \sigma^2}{G M / R} \simeq 1.
\end{equation}
This is slightly above the linewidth-size relation $\sigma \simeq 0.7 (R / \rm{1~ pc})^{1/2}~\rm{km~s^{-1}}$ \citep{Solomon1987, Heyer2004}, and is equivalent to $\sigma = 1.2 (R / \rm{1~ pc})^{1/2}~\rm{km~s^{-1}}$, which is well within the observed range (e.g. \citealt{Falgarone2009})

After driving for three cloud crossing times, we then turn off driving, turn on gravity and follow the subsequent gravitational collapse for approximately one global free fall time:
\begin{equation}
t_{\rm{ff}} = \sqrt{\frac{3 \pi}{32 G \bar{\rho}}} = 0.315~\rm{Myr},
\end{equation}
where $\bar{\rho}$ is the mean density of the box.  The simulations with radiation become prohibitively computationally expensive at late times and are stopped at $t = 0.83~t_{\rm{ff}}$ with a total stellar mass of 30 $M_{\odot}$ for simulation R.  The barotropic simulations are continued to $t = 1.05~t_{\rm{ff}}$ before they are stopped.  At this time the total stellar mass in simulation B is 50 $M_{\odot}$ compared to the total simulation mass of 185 $M_{\odot}$.  There is still gas bound to protostars totaling 11 $M_{\odot}$ when the simulations end.  Our stellar mass estimates may therefore be too low by 20\%.

Given our temperature of 10 K, the Jeans length at $\bar{\rho}$ is
\begin{equation}\label{eqn:jeanslengthdef}
\lambda_J = \left(\frac{\pi c_s^2}{G \bar{\rho}}\right)^{1/2} = 0.20~\rm{pc},
\end{equation}
and the Jeans mass is
\begin{equation}\label{eqn:jeansmassdef}
M_J = \frac{4 \pi}{3} \left(\frac{\lambda_J}{2}\right)^3 \bar{\rho} = 2.7~M_{\odot}.
\end{equation}
The turbulent Jeans mass, at density $\rho = \mathcal{M}^2 \bar{\rho}$, is 0.4 $M_{\odot}$.

The calculations have a $256^3$ base grid with 4 levels of refinement by factors of 2, giving an effective resolution of $4096^3$.  This resolution corresponds to $\Delta x_4 = 32$ AU at the finest refinement level.

\subsection{Evolution Equations}

We use the parallel adaptive mesh refinement code ORION for our simulations.
The numerical method is nearly identical to what we have used in previous papers \citep{Krumholz2007a, Offner2009, Cunningham2011, Krumholz2011, Myers2011}.  ORION solves the equations of compressible gas dynamics including self-gravity, radiative transfer, protostellar outflows, and radiating star particles, all on an adaptive grid.  Every cell in the grid has four conserved quantities: mass density, $\rho$, vector momentum density, $\rho \mathbf{v}$, gas energy density, $\rho e$, and radiation energy density, $E$.  These conserved quantities can be used to calculate derived quantities such as velocity, $\mathbf{v}$, and pressure, $P$.  In addition to the gas quantities, we evolve point-mass star particles, each with a position $\mathbf{x}_i$, mass $M_i$, momentum $\mathbf{p}_i$, angular momentum, $\mathbf{j}_i$ and luminosity $L_i$.  The subscript $i$ refers to the star particle number.  The particle method is explained in \citet{Krumholz2004} (hereafter KKM04), with the addition of radiation \citep{Krumholz2007a} and outflows \citep{Cunningham2011}.  The full set of evolution equations for gas and particles is

\begin{equation}\label{eqn:mass}
\frac{\partial \rho}{\partial t} + \nabla \cdot (\rho \mathbf{v}) + \sum_i [\dot{M}_{KKM04} W(\mathbf{r}_i) - \dot{M}_{w,i} W_w(\mathbf{r}_i) \xi(\theta_i)]= 0,
\end{equation}
\begin{eqnarray}
\frac{\partial \rho \mathbf{v}}{\partial t} + \nabla \cdot (\rho \mathbf{vv}) & = &- \nabla P - \rho \nabla \phi - \sum_i (\dot{\mathbf{p}}W(\mathbf{r}_i) - \label{eqn:momentum} \\
& &\dot{M}_{w,i} v_{w,i} W_w(\mathbf{r}_i) \xi(\theta_i) \cdot \hat{\mathbf{r}}_i), \nonumber
\end{eqnarray}
\begin{eqnarray}
\frac{\partial (\rho e)}{\partial t} + \nabla \cdot  [(\rho e + P)\mathbf{v}] & = & \rho \mathbf{v} \nabla \phi - \kappa_R \rho (4 \pi B - c E) - \nonumber\\
& & \left( \frac{\rho}{\mu m_H} \right)^2 \Lambda(T_g) - \label{eqn:gasenergy} \\
& & \sum_i [\dot{\varepsilon}_{KKM04} W(\mathbf{r}_i) - \nonumber \\
& & \dot{M}_{w,i} W_w(\mathbf{r}_i) \xi(\theta_i) \frac{k_B T_w \rm{K}}{\mu (\gamma - 1)}], \nonumber
\end{eqnarray}
\begin{eqnarray}
\frac{\partial}{\partial t} E - \nabla \cdot \left( \frac{c \lambda}{\kappa_R \rho} \nabla E \right) &=&  \kappa_P \rho \left(4 \pi B - c E \right) + \label{eqn:radenergy}\\
& &\left( \frac{\rho}{\mu m_H} \right)^2 \Lambda(T_g) + \sum_i L_i W(\mathbf{r}_i) \nonumber,
\end{eqnarray}
\begin{equation}\label{eqn:poisson}
\nabla^2 \phi = - 4 \pi G[\rho + \sum_i M_i \delta(\mathbf{r}_i)],
\end{equation}
\begin{equation}\label{eqn:sinkmassev}
\dot{M}_i = \frac{1}{1+f_w} \dot{M}_{KKM04},
\end{equation}
\begin{equation}\label{eqn:windmassev}
\dot{M}_{w,i} = f_w \dot{M}_i = \frac{f_w}{1+f_w} \dot{M}_{KKM04},
\end{equation}
\begin{equation}\label{eqn:sinkmomev}
\dot{\mathbf{p}}_i = \dot{\mathbf{p}}_{KKM04},
\end{equation}
\begin{equation}
\mathbf{r}_i = \mathbf{x} - \mathbf{x}_i,
\end{equation}
\begin{equation}
\theta_i = \rm{acos} (\hat{\mathbf{r}}_i \cdot \hat{\mathbf{j}}_i).
\end{equation}

The quantities entering these equations are defined below.  Equations (\ref{eqn:mass}) and (\ref{eqn:momentum}) are the fluid equations for mass and momentum, modified to include particles.  Equations (\ref{eqn:gasenergy}) and (\ref{eqn:radenergy}) are the energy equations for gas and radiation respectively. The Poisson equation for the gravitational potential, $\phi$ is given by equation (\ref{eqn:poisson}). The particle evolution is given by equations (\ref{eqn:sinkmassev}), (\ref{eqn:windmassev}) and (\ref{eqn:sinkmomev}).  We use periodic boundary conditions for all gas and particle quantities.

For the radiative runs, we adopt Marshak boundary conditions for the radiation field.  This allows radiation to escape from the box as it would from a molecular cloud.  The equation of state for the gas is given by
\begin{equation}
P = \frac{\rho k_B T_g}{\mu m_H} = (\gamma - 1) \rho \left(e - \frac{v^2}{2}\right),
\end{equation}
where $\mu = 2.33$ is the mean molecular weight for molecular gas of Solar composition and $\gamma$ is the ratio of specific heats.  Since most of the simulation domain is too cold to rotationally excite molecular hydrogen, we adopt $\gamma = 5/3$, representing a monatomic ideal gas.  The term $\kappa_P \rho \left(4 \pi B - c E \right)$ in equations (\ref{eqn:gasenergy}) and (\ref{eqn:radenergy}) represents energy exchanged between the radiation field and the dust in our gas, with $B = c a T_g^4 / 4 \pi$ representing the Planck emission function integrated over all frequencies.  The opacities $\kappa_P$ and $\kappa_R$ are Planck and Rosseland means given by the dust opacities from the iron normal, composite aggregates dust model of \citet{Semenov2003}.  We assume that the gas and the dust are thermally coupled.  When the gas temperature exceeds the dust destruction temperature, the energy exchange term goes to zero and the gas and radiation unrealistically decouple.  To address cooling from gas above the dust destruction temperature, we use the line cooling function $\Lambda(T_g)$ from \citet{Cunningham2006}.  This removes energy from the gas and adds that energy to the radiation field (see \citet{Cunningham2011} for further details).  The radiation flux limiter is given by $\lambda = \frac{1}{R} \left(\rm{coth} \mathit{R} - \frac{1}{\mathit{R}} \right)$, where $R = |\nabla E| / \kappa_R \rho E$ \citep{Levermore1981}.  It should be noted that we have excluded the radiation pressure and radiation enthalpy advection terms from equations (\ref{eqn:momentum}), (\ref{eqn:gasenergy}) and (\ref{eqn:radenergy}) that appear in the analogous equation in \citet{Krumholz2007a}.  This approximation is justified in the formation of low-mass stars, as shown in \citet{Offner2009}.

When radiation is neglected, the energy exchange term from equation (\ref{eqn:gasenergy}) disappears, and we close the system of equations with a barotropic equation of state for the gas pressure:
\begin{equation}
P = \rho c_{s0}^2 \left[1 + \left(\frac{\rho}{\rho_c}\right)^{\gamma-1}\right],
\end{equation}
where $c_{s0} = \sqrt{k_B T_0 / \mu m_H}$ is the isothermal sound speed at temperature $T_0 = 10$ K, $\gamma = 5/3$ and $\rho_c$ is the critical density.  The critical density determines the transition from isothermal to adiabatic regimes and we adopt $\rho_c = 2 \times 10^{-13}~\rm{g}~\rm{cm}^{-3}$ to agree with the collapse solution from \citet{Masunaga1998} prior to $\rm{H}_2$ dissociation.  Simulations that use the barotropic equation of state achieve maximum densities of $5 \times 10^{-15}~\rm{g~cm^{-3}}$, with an effective temperature at $\rho_{\rm{max}}$ of 10.8 K.  Most of the gas in any given simulations is effectively isothermal.

The particle quantities $\dot{M}_{KKM04}$, $\dot{\mathbf{p}}_{KKM04}$ and $\dot{\varepsilon}_{KKM04}$ in Equations (\ref{eqn:mass} - \ref{eqn:gasenergy}) represent the sink particle accretion rates of mass, momentum and energy from the surrounding gas in the absence of winds as given by KKM04.  The function $W$ represents a window function over the 4-cell accretion zone of the particle and $W_w$ represents the outflow window function.  Outflows are implemented as in \citet{Cunningham2011}, and summarized here.

Our outflow model is specified by the dimensionless parameter $f_w$, which sets the mass flux of outflow as a fraction of the accretion onto a star, and $v_{w,i}$, the wind launch speed.  The mass fraction in our simulations is $f_w = 0.3$. The wind speed is set by the Keplerian speed at the surface of the star, $v_{k,i} = \sqrt{G M_i / r_{*,i}}$ where $r_{*,i}$ is the protostellar radius, but is is capped at 60 km $\rm{s}^{-1}$ for computational speed.  Specifically, $v_{w,i} = \min(v_{k,i}, 60~\rm{km~s^{-1}})$.  The velocity cap has a similar effect to the choice of \citet{Cunningham2011} to use $v_{w,i} = v_{k,i} / 3$ for the most massive stars in the calculation.  Wind gas is injected at temperature $T_w = 10^4$ K.

The wind is injected over a window function $W_w$,
\begin{equation}
W_w = \frac{1}{C_1}\left\{
	\begin{array}{ll}
		\displaystyle r^{-2}   & \mbox{if } 4 \Delta x < r \leq 8 \Delta x \\
		0 & \rm{otherwise}
	\end{array}
\right..
\end{equation}
which represents a shell just outside of the accretion region.  The normalization constant $C_1$ is computed numerically to avoid numerical aliasing effects that occur from injecting a spherical wind into a Cartesian grid.

The exact angular distribution of the wind is described in the function $\xi$.  The functional form is taken from \citet{Matzner1999},
\begin{equation}\label{eq:windangle}
\xi(\theta) = \left[ \ln \left(\frac{2}{\theta_0} \right) ( \sin^2 \theta + \theta_0^2)\right]^{-1},
\end{equation}
where $\theta_0$ is a flattening parameter that sets the opening angle of the wind.  We use the fiducial value of $\theta_0 = 0.01$.  Equation (\ref{eq:windangle}) is averaged over the solid angle subtended by a grid cell at the outer radius of $W_w$.  This averaging is particularly important near $\theta = 0$.  In addition, $\xi$ is set to zero for $\theta \sim \pi/2$, so that winds are not injected directly into the plane of any equatorial disks.

We update the luminosity of each star, $L_i$, using the protostellar evolution model described in \citet{Offner2009}.  In this model, 75\% of the accretion energy is radiated away while 25\% is nominally used to power winds.  The energy of winds in our simulations is already determined by $f_w$ and $v_{w,i}$, which is typically 15\% of the accretion energy.  The remaining 10\% of the accretion energy is effectively lost.  When winds are not present, we still only radiate 75\% of the accretion energy for consistency across simulations.

The evolution equations can be described as the fluid and radiation equations from \citet{Offner2009} combined with the particle equations and line cooling of \citet{Cunningham2011}, but there is one important modification.  In the KKM04 sink particle methodology, all particles with overlapping accretion zones are merged together.  This gives an effective merger radius of 8 cells, or 256 AU at a grid resolution of 32 AU.  To limit this effect, we changed the merger radius to 4 cells, representing the point when a particle is in the accretion zone of another particle. This gives an effective merger radius of 128 AU at our resolution.  Even with this improvement, our particle algorithm will unrealistically merge stars that pass within 128 AU.  To address this, we have implemented a mass limit of $m_{\rm{merge}} = 0.05~M_{\odot}$, above which stars do not merge.  This limit is chosen to correspond to the mass of second collapse in the formation of a star with final mass 1 $M_{\odot}$\citep{Masunaga1998, Masunaga2000}.  Second collapse occurs when the protostar's core temperature becomes high enough to dissociate molecular hydrogen.  Before second collapse, protostars are extended balls of gas with radii of a few AU and have a much higher collisional cross-section than main sequence stars.  After second collapse, stellar mergers should be extremely rare and we do not allow them.  This approach is also used in \citet{Myers2011}.  The mass of second collapse depends on the accretion history of the protostar and is necessarily less than $0.05~M_{\odot}$ for brown dwarfs \citep{Stamatellos2009, Bate2011}.  The effects of numerical merger suppression is explored in \S 4.9.

ORION utilizes a second order Godunov scheme to solve the equations of compressible gas dynamics \citep{Truelove1998, Klein1999}.  These are equations (\ref{eqn:mass})-(\ref{eqn:gasenergy}), excluding terms from stars and radiation.  The Poisson equation (\ref{eqn:poisson}) is solved using a multi-grid iteration scheme \citep{Truelove1998, Klein1999, Fisher2002}.  The flux-limited diffusion radiation equation (\ref{eqn:radenergy}) and the radiation terms in equation (\ref{eqn:gasenergy}) are solved using the conservative update scheme from \citet{Krumholz2007b} modified to include the pseudo-transient continuation of \citet{Shestakov2008}.

We use the Truelove criterion \citep{Truelove1997} to determine the addition of new AMR grids so that the gas density in the calculation always satisfies
\begin{equation}
\rho < \frac{J^2 \pi c_s^2}{G(\Delta x_l)^2},
\end{equation}
where $\Delta x_l$ is the cell size on level $l$.  We adopt a Jeans number of 0.125.  In the simulations with radiative transfer, it is necessary to resolve the spatial gradients in the radiation field.  Areas of high radiation gradients are near accreting stars, which tend to already be refined under the Truelove criterion.  This is not always true for more evolved stars, which have higher luminosities and have accreted the dense gas that would trigger refinement.  We find that the radiation gradients are adequately resolved by refining whenever $|\nabla E| \nabla x_l / E > 0.25$.

\section{Results}

\subsection{Large Scale Evolution}

\begin{figure}[!htp]
\center{\includegraphics[scale = 0.6]{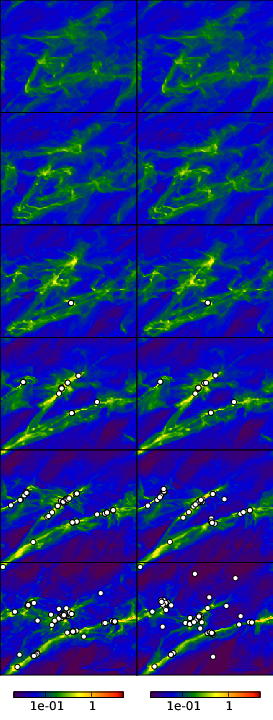}}\caption{Column density of the entire simulation domain for BW (left) and B (right) at times 0, 0.2, 0.4, 0.6, 0.8 and 1.0 $t_{\rm{ff}}$ from top to bottom.  Star particles are marked with white circles.  There is very little difference on the domain scale with and without winds for the barotropic simulations.  The color bar is $\rm{g~cm^{-2}}$ and the entire domain is 0.67 pc across.} \label{fig:nrtimage}
\end{figure}

\begin{figure}[!htp]
\center{\includegraphics[scale = 0.6]{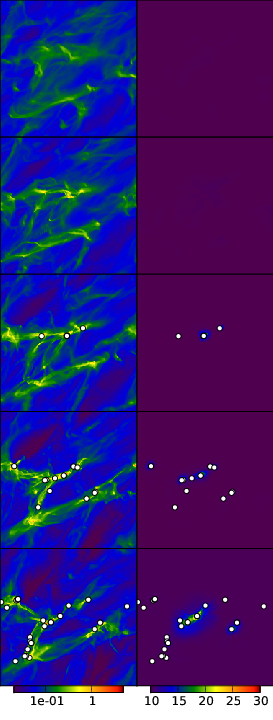}}\caption{Column density (left) and density weighted temperature (right) for simulation R at times 0, 0.2, 0.4, 0.6, and 0.8 $t_{\rm{ff}}$ from top to bottom. Star particles are marked with white circles.}\label{fig:rtnowindimage}
\end{figure}

\begin{figure}[!htp]
\center{\includegraphics[scale = 0.6]{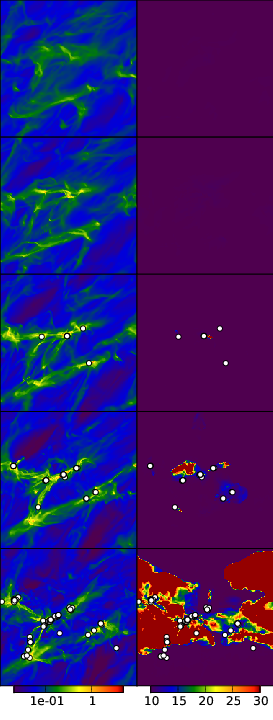}}\caption{Same as Figure \ref{fig:rtnowindimage}, but for simulation RW. The high temperature regions are the paths of outflows.  It only takes a small amount of gas at $10^4$ K to move the average temperature above 30 K through that line of sight.} \label{fig:rtimage}
\end{figure}

The evolution of the barotropic simulations with and without winds is depicted in Figure \ref{fig:nrtimage}.  Figures \ref{fig:rtnowindimage} and \ref{fig:rtimage} depict the evolution of the radiative simulations without and with winds, respectively.
In all simulations, for $t \lesssim 0.4 t_{\rm{ff}}$, there are cloud-scale filaments that slowly contract, allowing $3$ turbulent cores of width $\sim20,000$ AU and density $\sim 10^{-19}~\rm{g~cm}^{-3}$ to form.  This length is half the Jeans length at the average cloud density.  The first core to form is at the center of each panel in Figures \ref{fig:nrtimage}-\ref{fig:rtimage}, the second core is left of the central core, near the left edge, and the third core is near the right edge, at the end of a simulation-wide filament.  At this point, the cores begin to fragment, while new cores form, eventually forming 6 fully developed cores.  These cores each have a central stellar system.  In simulation R, this central system is all that forms in each core.  In all other simulations, the central system represents $\sim75\%$ of the stellar mass in the core and an additional group of low-mass stars forms, totaling $\sim10$ stars per core.  These cores with multiple stars generally resemble observed high-stellar-density cores \citep{Teixeira2007, Chen2010}.  There are an additional 20 stars in cores that are still forming at the end of the simulation, giving a global total of 80 stars.  Three of the cores coalesce by the end of the simulation to form a single group of 30 stars.

\begin{figure}[!htp]
\center{\includegraphics[width=3.5in]{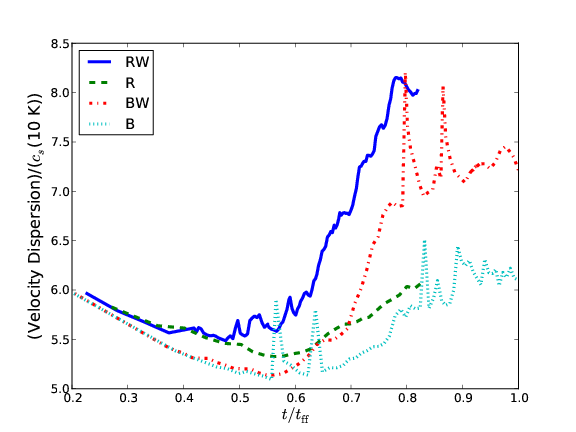}}\caption{Time evolution of global rms Mach number for simulations with and without winds and with and without radiation.  The turbulent energy in all simulations decays for half a global free fall time, at which point gravitational potential energy from stars is converted into kinetic energy, which raises the rms velocity.  When winds are included, they contribute over twice as much energy as gravity itself.} \label{fig:machvt}
\end{figure}

The evolution of the 3D rms velocity dispersion, $\sigma_v$, is shown in Figure ~\ref{fig:machvt}.
The global turbulence decays until star formation ramps up at $t \sim 0.5 t_{\rm{ff}}$.  There are two main mechanisms for star formation to increase $\sigma_v$.  First, as stars accrete mass and deepen their gravitational potential, the surrounding gas can convert gravitational energy into kinetic energy as it falls into the stars.  This is shown in the gradual increase in Mach number for $t > 0.6 t_{\rm{ff}}$ in the B simulation.  This effect is strong enough to return $\sigma_v$ to near its original virialized value by itself.  In rare cases, a many-body close encounter between stars will eject some gas at high velocities.  There is not much momentum injected this way and the energy quickly dissipates, but it causes spikes in $\sigma_v$ for the barotropic simulations, which have more small-scale fragmentation and therefore more many-body close encounters.  The second mechanism occurs when protostellar winds are included.  Some mass accreted onto stars is directly injected around the stars at high velocities.  This causes the smooth increase in $\sigma_v$ for simulations BW and RW as well as spikes from events with particularly high accretion rates that lead to bursts in wind momentum.

\begin{figure}[!ht]
\center{\includegraphics[width=3.5in]{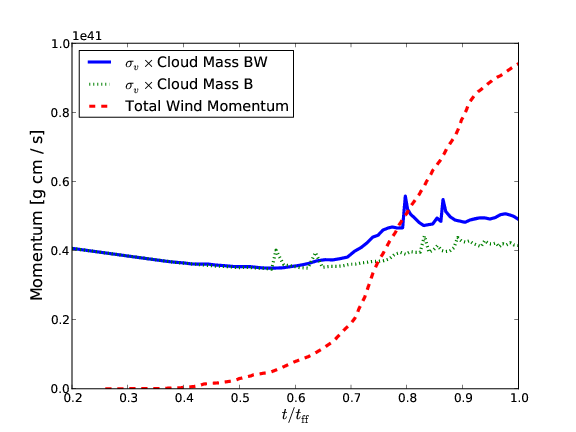}}\caption{Total momentum that has been injected by winds over time for the barotropic simulations.  For comparison, the total mass of the simulation multiplied by the velocity dispersion is also plotted.  The total wind momentum integrates all injected momentum over time, even from winds that have decayed.  As a result, the amount of injected momentum is eventually higher than the actual momentum of the cloud.} \label{fig:momvt}
\end{figure}

The total momentum injected by winds for model BW is shown in Figure \ref{fig:momvt}.
For comparison, a characteristic value of the magnitude of the momentum associated with internal motions in the cloud, $M_{\rm{cloud}} \sigma_v$ is also plotted.  For $t > 0.8 t_{\rm{ff}}$, the total momentum that has been injected by the winds is greater than the characteristic cloud momentum.  At this point, the total amount of turbulent momentum that has been dissipated (including dissipation of wind momentum) is roughly the total amount of momentum that has been injected by the winds.  By the end of the BW and RW simulations, the total wind momentum injected into the cloud is over twice the characteristic cloud momentum.  The kinetic energy injected from the winds dissipates over time, suggesting a steady-state solution where the velocity dispersion of the cloud is constant with time as the winds replenish energy as quickly as it can dissipate.

\subsection{Evolution of the Protostellar Population}

\begin{figure}[!ht]
\center{\includegraphics[width=3.5in]{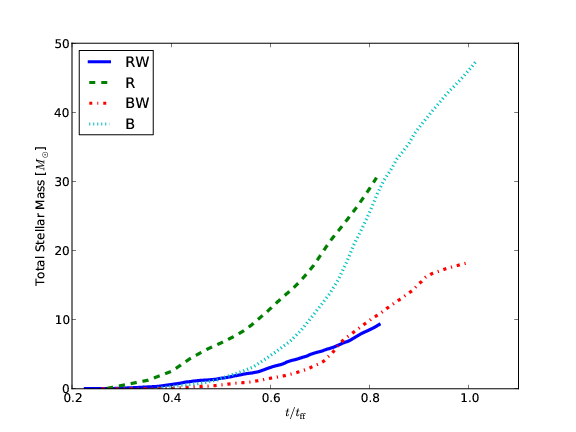}}\caption{Total mass in stars as a function of time for the four simulations.  The mass of the entire simulation domain is 180~$M_{\odot}$.} \label{fig:mvt}
\end{figure}

\begin{figure}[!ht]
\center{\includegraphics[width=3.5in]{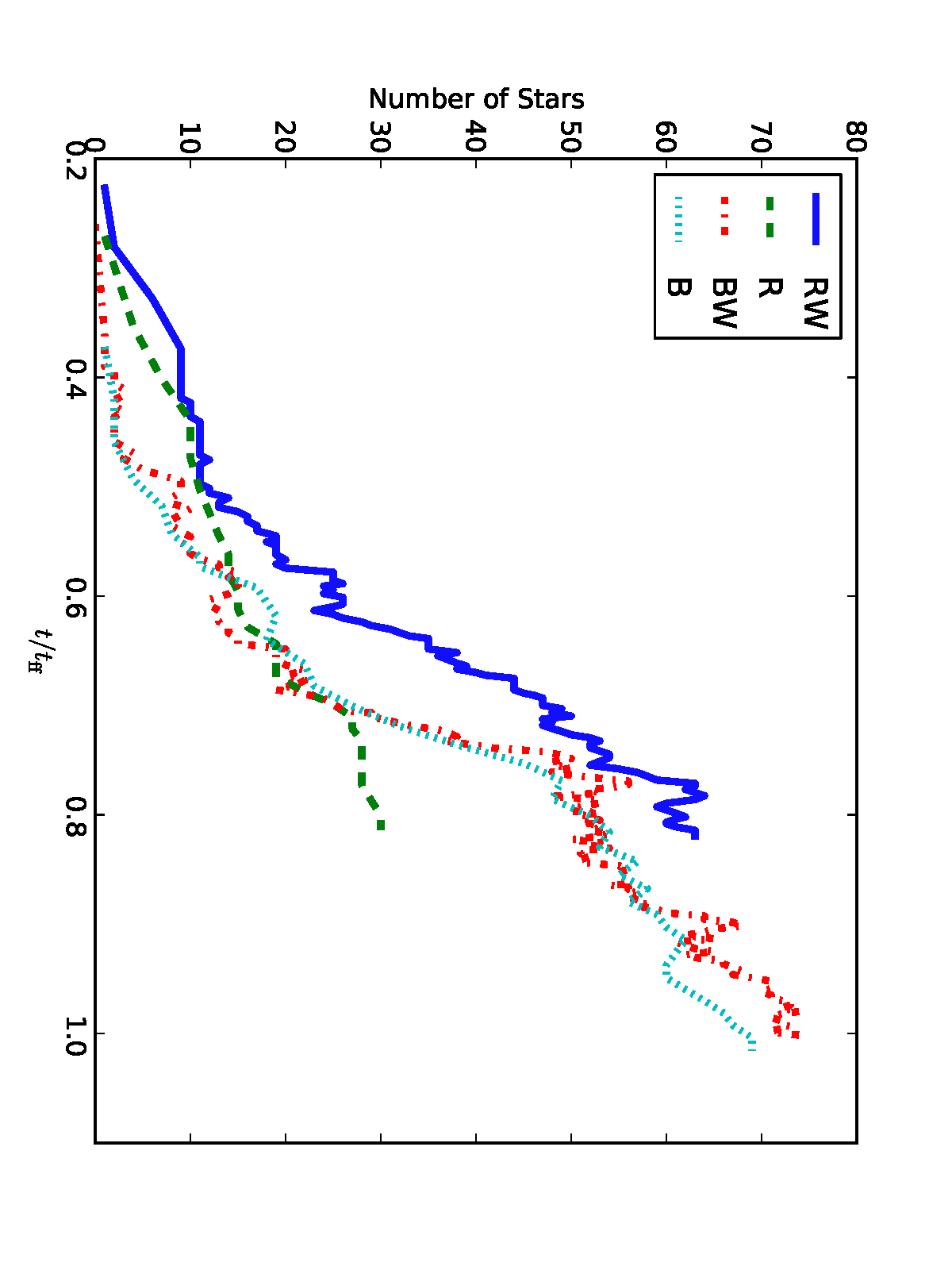}}\caption{Number of stars as a function of time for the four simulations.  In the barotropic case, the number of stars is unaffected by winds.  In the radiative case, radiation suppresses the number of stars unless winds are present.} \label{fig:nvt}
\end{figure}

The total mass in stars and the total number of stars as functions of time is shown are Figures \ref{fig:mvt} and \ref{fig:nvt}.  The realization of the initial turbulence is slightly different between the radiative and barotropic simulations, so star formation begins at different times.  This is the result of a slightly higher maximum density due to turbulence and it is unrelated to the choice of radiative or barotropic thermodynamics.  The first stars form at $t = 0.2 t_{\rm{ff}}$ for the radiative simulations and $t = 0.35 t_{\rm{ff}}$ for the barotropic simulations.  The turbulence overlaps enough between the radiative and barotropic cases, however, that at later times the total mass in stars at any given time is similar for the two cases when winds are not included.  Winds reduce the mass in stars by about a factor of 3 in both the radiative and barotropic cases.  The number of stars does not change between BW and B, implying winds do not cause or suppress fragmentation by themselves.  The number of stars in RW is significantly greater than in R, however, because protostellar luminosity inhibits fragmentation and the winds reduce that luminosity.  The three simulations other than R show a dramatic increase in the number of stars at $0.6 ~t_{\rm{ff}} < t < 0.8~t_{\rm{ff}}$.

\begin{figure}[!ht]
\center{\includegraphics[width=3.5in]{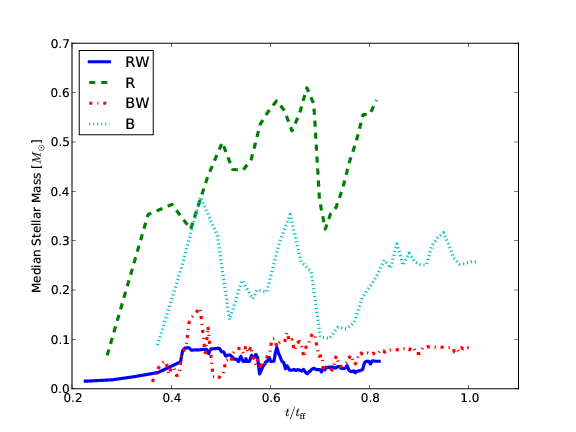}}\caption{Median mass of stars as a function of time for the four simulations.  The two cases with winds maintain low medians throughout the simulations.  The case with radiation without winds (case R) is able to suppress fragmentation and new star formation largely stops as the original stars accrete mass.} \label{fig:mmedvt}
\end{figure}

The evolution of median stellar mass is shown in Figure \ref{fig:mmedvt}.  This is a rough proxy for the characteristic mass of the protostellar mass function.  Note that it will always be lower than the median mass of the IMF, because not all of the stars have finished accreting.  Both BW and RW maintain a median around 0.05 $M_{\odot}$ (similar to that of the protostellar mass functions in \citealt{McKee2010}) throughout the simulation.  The median does increase for simulation BW around $t > t_{\rm{ff}}$ as the formation rate of new stars decreases.  The median of B fluctuates more, but is around 0.2 $M_{\odot}$.  Lastly, R maintains a median around $0.5 M_{\odot}$.  This general behavior should be expected.  The median mass is lowest when winds are included and highest when radiation is allowed to suppress fragmentation.  The case with both winds and radiation ends up similar to BW because winds reduce protostellar luminosities.

\begin{figure}[!ht]
\center{\includegraphics[width=3.5in]{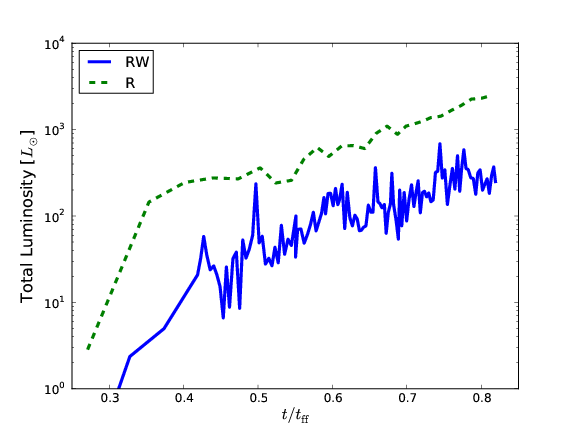}}\caption{Total stellar luminosity versus time for simulations with radiation, both with and without winds.  Winds dramatically lower the luminosity.} \label{fig:Lvt}
\end{figure}

The global luminosity evolution for the radiative simulations
 is plotted in Figure \ref{fig:Lvt}.  The winds reduce the total luminosity by a factor of up to 10 at any given time.  This is expected since the global accretion luminosity is
\begin{equation}\label{eqn:Lacc}
L=\frac{3}{4}\ \sum_i \frac{G M_{\star i} \dot M_{\star i} }{ R_{\star i}} \sim \frac {G M_{\star,\rm tot} \dot M_{\star,\rm tot} }{ \langle R_*\rangle};
\end{equation}
given that the total mass in stars $M_{\star, \rm tot}$ is reduced by a factor of 3 when winds are included, and the total accretion rate of stars $\dot{M}_{\star, \rm tot}$ is also reduced by 3, the total luminosity is therefore reduced by a factor of 9 assuming the characteristic stellar radius, $\langle R_{\star} \rangle $, does not change.  Main sequence stars typically have a positive correlation between mass and radius, suggesting the factor of 9 is an upper limit.  However, at any given point in time, many stars in our simulations are in the degenerate regime where mass and radius are negatively correlated \citep{Chabrier2009}, which counteracts the positive correlation from the higher mass stars and keeps the total luminosity ratio near 9.

\begin{figure}[!ht]
\center{\includegraphics[width=3.5in]{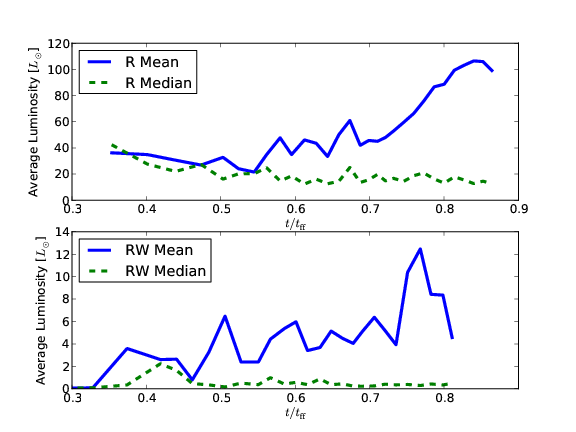}}\caption{Mean and median stellar luminosity versus time for simulations with radiation.  The top panel is the simulation without winds and the bottom panel is the simulation with winds.} \label{fig:Lavevt}
\end{figure}

The average stellar luminosity is shown in Figure \ref{fig:Lavevt}.  As was seen in the plot of total luminosity, the mean and median values of protostellar luminosity are much lower when winds are included.  The disparity in average luminosity is even greater than the disparity in total luminosity because there are fewer stars when winds are excluded.  The average luminosity in simulation R is heavily influenced by a single 6.6 $M_{\odot}$ star that accounts for over half the total luminosity in the simulation.  Unlike the low-mass stars, most of this luminosity is powered by nuclear fusion rather than accretion.  Protostellar luminosities will be discussed further in section \ref{sec:disc:PLF}

\subsection{Thermal Evolution}

\begin{figure}[!ht]
\center{\includegraphics[width=3.5in]{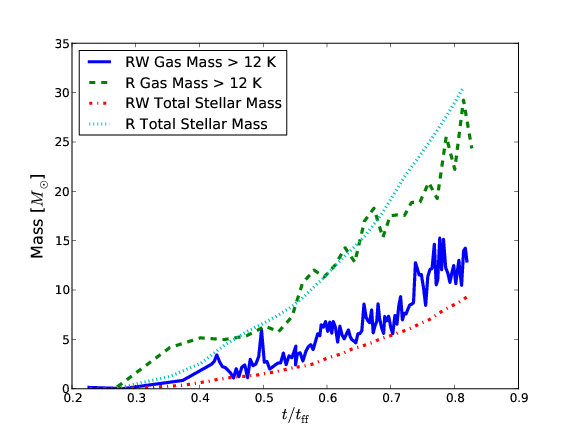}}\caption{Total gas mass heated above 12 K versus time, compared to the background value of 10 K.  The total mass in stars is also plotted for reference.} \label{fig:warmgasvt}
\end{figure}

All simulations start at a background temperature of 10 K and are bathed in 10 K radiation.  Stellar radiation and mechanical energy from protostellar outflows, can raise this temperature.  We have identified gas heated above 12 K as thermally affected by stellar feedback.  Turbulent dissipation alone heats almost no gas above 12 K, leaving stellar feedback as the only explanation for this heating.  The total mass of this gas is shown in Figure \ref{fig:warmgasvt}.
The simulation with winds has significantly less heated gas than the simulation without winds.  This is due to the reduced luminosity caused by the winds shown in Figure \ref{fig:Lvt}.  In each simulation, the mass in heated gas roughly follows the mass in stars.  When winds are included, the mass in stars drops by a factor of 3, and the mass of heated gas also falls due to the reduced stellar luminosity.  Some of the lost radiative heating is replaced by mechanical heating from outflows.  This is evident in the ratio of heated gas mass to stellar mass.  The mean value of this ratio is 0.9 without winds, but rises to 1.5 with them.  Wind gas is injected at a temperature of 10$^4$ K, but cools quickly.  The mass in gas with temperatures above 1000 K is less than 2\% of the mass in stars at any given time.

\begin{figure}[!htp]
\center{\includegraphics[width=3.0in]{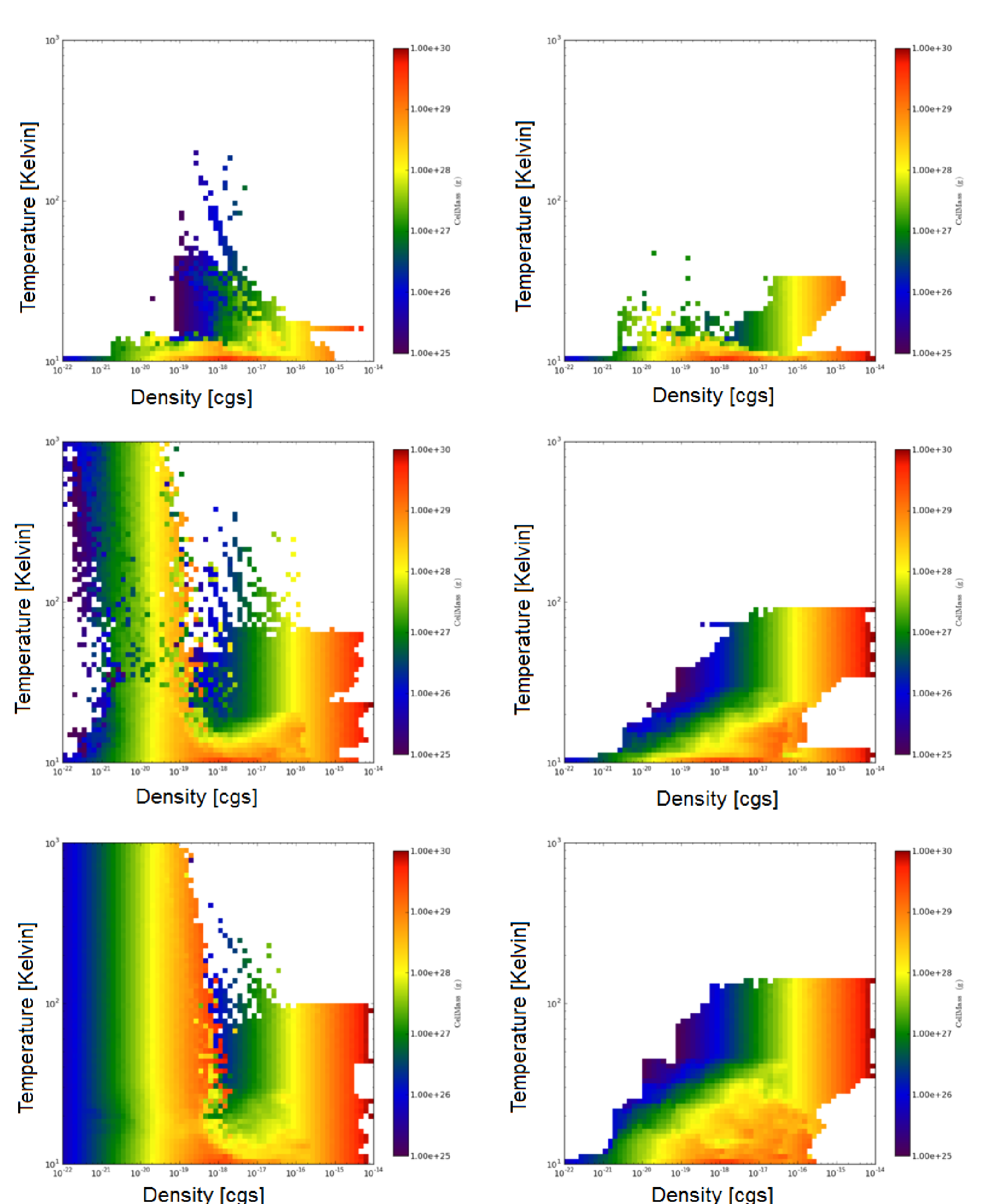}}\caption{Phase plot showing total gas mass as a function of temperature (y-axis) and density (x-axis) for radiative simulations with (left) and without (right) winds.  Phase plots are taken at times of 0.25, 0.5 and 0.75 $t_{\rm{ff}}$ from top to bottom.  The high-temperature, low-density gas on the left part of the wind phase plots is outflow gas.  Warm, high-density gas is gas near a luminous star.} \label{fig:phase}
\end{figure}

To further explore the heated gas, temperature-density phase plots with and without winds are shown in Figure \ref{fig:phase}.
The phase plots with and without winds are notably different in two areas.  First, the wind gas fills the high-temperature, low-density domain, while the same domain is empty without winds.  Second, high-density gas with $\rho > 10^{-16}~\rm{g/cm}^{3}$ has a higher temperature range without winds than with winds because the extra stellar luminosity heats that gas.  When winds are included, that dense gas is less common in addition to being colder; there is more fragmentation, which turns dense gas into stars.  In addition, some of the gas is also blown away by the winds themselves.

\section{Discussion}
\subsection{Supporting a Cloud with Outflows}
\begin{figure}[!ht]
\center{\includegraphics[width=3.5in]{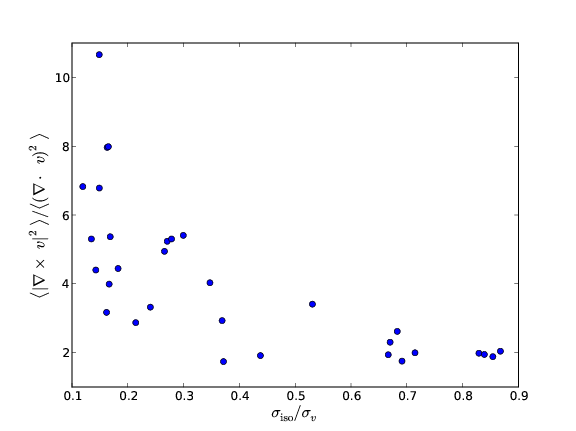}}\caption{Ratio of solenoidal to compressive velocities squared ($\langle | \nabla \times v |^2 \rangle / \langle (\nabla \cdot v) ^2 \rangle$) versus time with and without winds for the radiative simulations.  For pure hydrodynamic, isotropic turbulence, this ratio is around 2.  This ratio stays near 2 when winds are excluded.  When winds are included, the turbulence is much more anisotropic, leading to higher solenoidal fractions.} \label{fig:solevt}
\end{figure}

The turbulent evolution shown in Figures \ref{fig:machvt} and \ref{fig:momvt} roughly agrees with previous simulations of molecular clouds with outflows, such as those in \citet{Nakamura2007} and \citet{Wang2010}.  Turbulent energy decays initially, only to be replaced by kinetic energy from winds and from gravity.  While these sources can increase the total kinetic energy of gas, the new turbulence is fundamentally different from the isotropic, homogeneous hydrodynamic turbulence it replaces.  This result is also seen in \citet{Nakamura2007}, who find that the late time turbulent statistics do not match expected isotropic hydrodynamic results.  One key difference is the energy from outflows is highly anisotropic.  Outflow cavities are marked by long channels with high velocity shear between the fast outflow gas and the slow ambient gas.  This shear is detectable as solenoidal energy.  There is some compressive energy at the head of the outflow cavity, but most of the surface area is the side walls of the cavity and not the head.  To measure the relative importance of solenoidal and compressive motions, we use the ratio $\langle | \nabla \times v |^2 \rangle / \langle (\nabla \cdot v) ^2 \rangle$, shown in Figure \ref{fig:solevt}.  In the case of isotropic turbulence, this ratio is 2, which is also the ratio of solenoidal to compressive energies \citep{Elmegreen2004}.
Wind injection greatly increases the solenoidal velocities, steadily increasing the solenoidal to compressive ratio over the course of the simulation.  Bursts in outflows injection around $t = 0.5 t_{\rm{ff}}$ and $t = 0.65 t_{\rm{ff}}$ are also visible in the solenoidal velocity.  Anisotropic turbulence behaves differently than isotropic turbulence; in particular, it takes longer to decay since it decays on the crossing time calculated from the smallest velocity dispersion \citep{Hansen2011}.  It is difficult to measure this increase in the decay time in our simulations, however, because the winds drive turbulence over a wide range of scales.

\begin{figure}[!htp]
\center{\includegraphics[width=3.5in]{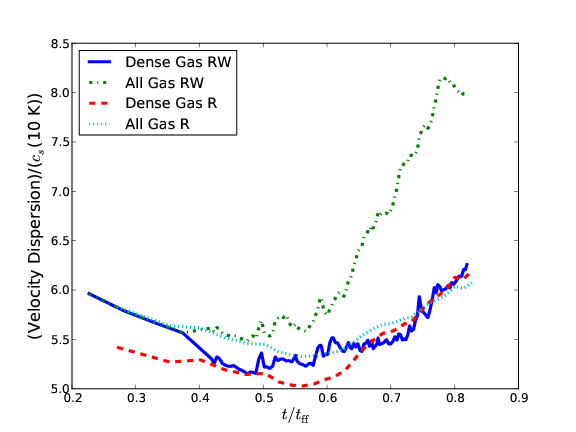}}\caption{Time evolution of rms Mach number of dense gas and all gas with and without winds.  Winds significantly raise the Mach number of the light gas, but do not strongly influence the dense gas turbulence.} \label{fig:machheavyvt}
\end{figure}

The other major difference between the initial turbulence and wind-driven turbulence is seen in the rms velocity, $\sigma_{\rm{dense}}$, of gas with $\rho > \bar{\rho}$.  Even in isotropic, homogenous, hydrodynamic turbulence, there is a negative correlation between density and velocity, causing $\sigma_{\rm{dense}} < \sigma_v$ \citep{Offner2009b}.  The winds themselves are collimated, very low density gas and have difficulty transmitting energy into high density gas. This means that while $\sigma_v$ is much greater with winds than without, $\sigma_{\rm{dense}}$ does not change much when winds are included.  The evolution of $\sigma_{\rm{dense}}$ compared to $\sigma_v$ is shown in Figure \ref{fig:machheavyvt}.

Because the dense gas is relatively unaffected by the outflows, if our cloud had been centrally concentrated like that of \citet{Nakamura2007} or \citet{Wang2010}, the dense part of the cloud would have most likely collapsed on itself even with the support of protostellar outflows.  Magnetic fields may have an effect, as shown by \citet{Wang2010}.  The magnetic fields help transmit outflow energy to a much larger solid angle, so that even a centrally concentrated cloud can achieve a quasi-static balance between outflows and gravity.

\subsection{Comparison to the IMF}
\begin{figure}[!htp]
\center{\includegraphics[width=3.5in]{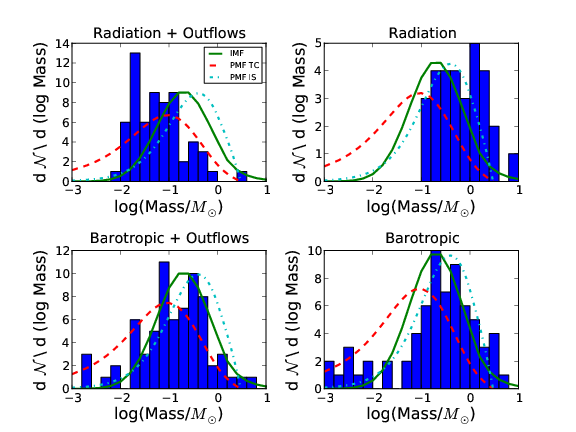}}\caption{The mass function of all stars in each simulation are shown in blue histograms.  The stellar IMF from \citet{Chabrier2005} is plotted as the solid green line.  The protostellar mass functions for the turbulent core model and the isothermal sphere model from \citet{McKee2010} are the dashed lines and dash dotted lines, respectively.  Top left:  RW at $t = 0.83 t_{\rm{ff}}$. Top right: R at $t = 0.83 t_{\rm{ff}}$. Bottom left:  BW at $t = 1.09 t_{\rm{ff}}$.  Bottom right:  B at $t = 1.03 t_{\rm{ff}}$} \label{fig:imf}
\end{figure}

Our simulations are marginally able to capture binaries, so most star particles should represent a single stellar object instead of a stellar system.  The typical binary separation for main sequence stars is 50 AU \citep{Mathieu1994}, just slightly larger than our resolution of 32 AU.  We cannot form stars within 128 AU of another star due to merging star particles, but stars that form beyond 128 AU can spiral in to 32 AU through interaction with gas.  At any given time, about 2/3 of our star particles are in stellar multiples.  The multiple properties are dynamic due to unstable high-order multiples and we cannot compare to the observed system properties.  We can, however, compare to observed stellar properties.  The mass functions of the four main simulations are shown in Figure \ref{fig:imf} and compared to the stellar IMF in \citet{Chabrier2005} as well as protostellar mass functions from \citet{McKee2010}.
The mass functions in Figure \ref{fig:imf} are shown at the latest time available for each simulation.  The barotropic simulations could be evolved to later times due to the computational expense of flux-limited diffusion with many stellar sources.  The mass functions in Figure \ref{fig:imf} are not exactly comparable to an IMF because some of the stars are still accreting.  The barotropic mass functions have evolved to a later time and should be similar to the IMF.  The radiative simulations are still actively accreting and are closer to the protostellar mass functions in \citet{McKee2010}.  Among the protostellar mass functions, the turbulent core and competitive accretion (not shown in Figure \ref{fig:imf}) models both roughly match RW.  The isothermal sphere protostellar mass function is the best fit to mass functions of the simulations without winds, but this is merely a function of both mass functions being too top-heavy.  The actual accretion is not described by an isothermal sphere, and is discussed further in section \ref{sec:disc:PLF}.

As expected, the simulations without winds have mass functions skewed to higher masses than those with winds.  The mass functions of both simulations without winds have too much mass at the high end compared to the IMF.  The best fit to the IMF is from the BW simulation.  It should be noted that the normalized mass functions for BW and RW look nearly identical when compared at the same time (explored in the next section).  It is a good assumption that RW would also eventually match Chabrier at $t \sim t_{\rm{ff}}$.

\subsection{Comparison to Protostellar Luminosities}\label{sec:disc:PLF}

Theoretical predictions of protostellar luminosities are often too high compared to observations of regions of low-mass star formation \citep{Kenyon1990, Young2005, Enoch2009}, so it is important to compare our own simulations with observations.  The mean and median luminosities of protostars observed in nearby clusters are $\langle L_{\rm{obs}} \rangle = 5.3^{+2.6}_{-1.9}~L_{\odot}$ and $L_{\rm{obs,med}} = 1.5^{+0.7}_{-0.4}~L_{\odot}$, respectively \citep{Enoch2009, Evans2009, Offner2011}.  The typical mean and median luminosity in our simulation with winds are $\langle L \rangle = 6.9~L_{\odot}$ and $L_{\rm{med}} = 1.4~L_{\odot}$, in agreement with the observations.  An additional useful quantity is the standard deviation of the luminosity, $\sigma(L)$.  This is sensitive to outliers in the luminosity distribution, but this can be mitigated by using the log of the luminosity.  We find $\sigma(\log L) = 0.77~\rm{dex}$.  This matches the observed value $\sigma(\log L_{\rm{obs}}) = 0.7^{+0.2}_{-0.1}$.  For these comparisons, we have discarded stars with $L < 0.05~L_{\odot}$, below the detection limit of the observations.

Theoretical frameworks that do match the observed protostellar luminosities are presented in \citet{McKee2010} and \citet{Offner2011}.  These models differ from the straightforward, isothermal sphere models \citep{Fletcher1994a, Fletcher1994b} in three important ways.  First, \citet{Offner2011} assumed that 1/4 of the energy of the gas that accreted onto the star was removed non-radiatively.  This effect is captured in our simulations.  Second, they considered accretion rates that rise over time, such as predicted in core accretion and competitive accretion.  These accretion rates all take the form,
\begin{equation}\label{eqn:mountapered}
\dot{m} = \dot{m_1} \left(\frac{m}{m_f}\right)^j m_f^{j_f},
\end{equation}
where $m$ is the instantaneous mass of a protostar, $m_f$ is the final mass of a protostar, and $\dot{m_1}$ is a constant throughout a cloud.  More realistic accretion rates will rise at early times and slowly decline over time, in what \citet{McKee2010} call `tapered accretion'.  There are multiple approaches to this; the one taken by \citet{McKee2010} is to assume a linear decrease in the accretion rate with time, $\dot{m} \propto 1-t/t_f$, which implies that
\begin{equation}\label{eqn:motapered}
\dot{m} = \dot{m_1} \left(\frac{m}{m_f}\right)^j m_f^{j_f}\left[1 - \left(\frac{m}{m_f}\right)^{1-j}\right]^{1/2}.
\end{equation}
The values of $j$ and $j_f$ for several different accretion theories are shown in table \ref{tab:acc} as well as the values measured in our simulations with winds.  Radiative and barotropic simulations produce the same $j$ and $j_f$.  The fits shown are for equation (\ref{eqn:motapered}).  Our data do not agree with the functional form for untapered accretion (equation (\ref{eqn:mountapered})), but if this were used, $j_f$ would remain the same while $j \sim -0.1$.
\begin{table}[!htp]
\begin{flushleft}
\caption{Accretion rate dependencies on instantaneous and final protostellar mass}\label{tab:acc}
\begin{tabular}{|c|c|c|}

  \hline
  Accretion Mechanism$^a$ & $j$ & $j_f$ \\
  \hline
  Isothermal Sphere & 0 & 0 \\
  Turbulent Core Accretion & 0.5 & 0.75 \\
  Competitive Accretion & 0.67 & 1\\
  Simulations with Winds & $0.3 \pm 0.2$ & $0.6 \pm 0.2$\\

  \hline

\end{tabular}
\end{flushleft}
$^a$\footnotesize{All accretion is tapered, as in equation \ref{eqn:motapered}}
\end{table}
Our measured values of $j = 0.3 \pm 0.2$ and $j_f = 0.6 \pm 0.2$ marginally agree with the turbulent core accretion model, but do not match any of the other theories.  Our $j$ and $j_f$ actually lie in between isothermal sphere and turbulent core accretion.  This suggests the most appropriate theories may be the the two component turbulent core (2CTC) model, which is approximately isothermal sphere accretion at early times and turbulent core accretion at later times.  Other candidates include the TNT model \citep{PhilMyers1992}, which includes both thermal and nonthermal support, and the core-clump accretion model from \citet{PhilMyers2011b} which is approximately isothermal sphere accretion at early times and reduced Bondi accretion at late times.  Our low $j$ does not agree with competitive accretion, which accelerates with mass.

While we agree with tapered accretion from \citet{McKee2010}, we do not rule out other tapered models, such as the exponential tapering in \citet{Schmeja2004}.  Our average accretion rates are lower than \citet{Schmeja2004} by an order of magnitude, partially due to our inclusion of protostellar outflows, but we see a similar rise and fall of accretion, which can be fit by many functional forms.

The last effect considered by \citet{McKee2010} and \citet{Offner2011} is episodic accretion from FU Ori type protostars (e.g. \citealt{Hartmann1996}).  If protostars spend most of their life in a low-accretion, low-luminosity phase and accrete nearly all their mass during short intense accretion bursts, the median stellar luminosity can be greatly reduced.  This episodic accretion is thought to arise from disk instabilities \citep{Basu2005}.  \citet{Offner2011} estimated that 25\% of the mass was accreted in this manner.  We do not resolve disks and therefore do not see episodic accretion in our simulations.  Given that our simulations match the observed luminosity function without episodic accretion, models that rely on substantial episodic accretion \citep{Dunham2010, Stamatellos2011} may no longer be necessary.

\subsection{Suppressing Fragmentation with Radiative Feedback}
Fragmentation in the simulations occurs in two distinct phases.  First, the cloud as a whole forms cores of size $\sim20,000$ AU (0.1 pc).  This size scale is half the Jeans length at density $\bar{\rho}$.  These large scale cores each have a major filament within them that is above the critical line density for stability \citep{Larson1985, Inutsuka1992, Inutsuka1997},
\begin{equation}
\lambda_{\rm{crit}}  = \frac{2 k T}{\mu m_H G}.
\end{equation}
At our temperature, $T = 10$ K, the critical line density is $\lambda_{\rm{crit}} = 1.0 \times 10^{16}~\rm{g~cm^{-1}}$.  The line densities of the filaments in our cores range from $1.7\times10^{16}$ to $4.0 \times 10^{16}~\rm{g~cm^{-1}}$, similar to filament line densities seen in Serpens \citep{Andre2010}.  The general morphology is similar to the hub-filament structure in \citet{PhilMyers2009} and \citet{PhilMyers2011} with the hub at the center of each core.

The large scale fragmentation is not affected by radiation.  In contrast to high-mass star formation \citep{Krumholz2007a, Cunningham2011, Krumholz2011}, there is simply not enough protostellar luminosity to affect the large scales except possibly at late times in the R simulation.  The winds do travel through the entire simulation domain and could theoretically affect the fragmentation, but this does not happen in practice since the winds do not couple well to the cores.

The second stage of fragmentation occurs as the filament in each core contracts under self-gravity.  The filaments can then fragment into many stars around the central stellar system.  Unlike the large-scale core fragmentation, this small scale fragmentation can be significantly suppressed by radiation \citep{Krumholz2007a, Offner2009}.  This is best demonstrated in Figure \ref{fig:nvt}.  Simulation R stagnates at 30 stars, while B and BW finish with 80 stars.  At the end of R, the total protostellar luminosity is 2500 $L_{\odot}$ and this is currently heating 30 $M_{\odot}$ of gas.  Winds by themselves do not affect small scale fragmentation.  The total number of fragments, and therefore the total number of stars, are the same in the barotropic simulations with and without winds, as shown in Figure \ref{fig:nvt}.  It should be noted that our fragments occur in the core and not in accretion disks.  This agrees with higher resolution core evolution simulations \citep{Offner2010}.

\begin{figure}[!ht]
\center{\includegraphics[width=3.5in]{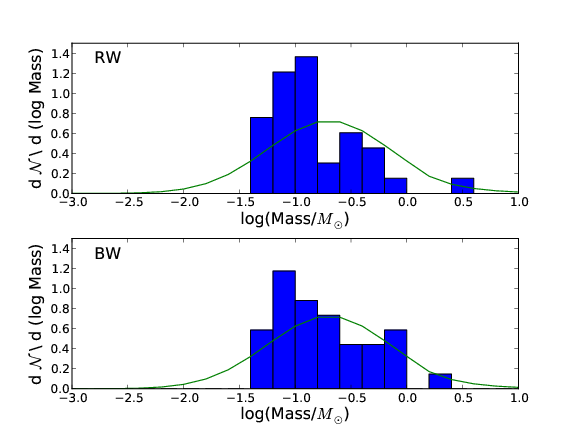}}\caption{Mass function for all stars with mass $> 0.05 M_{\odot}$ for simulations RW (top) and BW (bottom).  The Chabrier IMF is plotted to guide the eye, but the simulations are still accreting and are not expected to match the IMF.  The radiative and barotropic mass functions are similar, showing that radiation does not significantly affect the IMF in regions of low-mass star formation when winds are present.  The main difference comes from the largest star in RW fragmenting more in BW, lowering the largest mass and adding more stars around $0.5 M_{\odot}$.} \label{fig:imfRWvBW}
\end{figure}

When winds and radiation are combined, the winds have a significant indirect effect.  The winds lower the mass and accretion rate of the protostars, which lowers the luminosity as seen in Figure \ref{fig:Lvt}.  This means that the fragmentation suppression seen in comparing R to B should be reduced when comparing RW to BW.  To investigate this in more detail, we have shown the mass functions of RW and BW, both at $t = 0.83 t_{\rm{ff}}$, in Figure \ref{fig:imfRWvBW}.  For purposes of comparison, we have excluded stars with mass $M < 0.05 M_{\odot}$, since these stars are usually short lived and possibly subject to details of numerical sink particle formation or merging.
The two mass functions are nearly the same.  A two sample Kolmogorov-Smirnov test gives a 50\% chance that both samples are drawn from the same underlying function.  The primary difference between the two mass functions lies in the heaviest star in the simulation.  The core that forms the heaviest star fragments early in BW, turning a star that is 2.8 $M_{\odot}$ in RW into a 2.0 $M_{\odot}$ star in BW with two extra $M \sim 0.4 M_{\odot}$ stars.  The reduced fragmentation in RW for that star might be expected from radiative feedback, since that is the most luminous star in the simulation.

To understand why radiation does not significantly alter star formation for regions
of low-mass star formation such as we are considering, we expand on a line of reasoning
developed by \citet{Bate2009}. He introduced an effective Jeans length, which we write
as $\lambda_\eff$, defined by the condition
\beq
\lambda_\eff\equiv \frac{GM}{c_s^2},
\eeq
so that $\lambda_\eff$ is the radius at which the escape velocity equals the isothermal
sound  speed, $c_s$. Since the mass inside $\leff$ is $M=(4\pi/3)\rho \lambda_\eff^3$,
one finds that Bates' effective Jeans length is smaller than the standard
Jeans length, $\lj$, which is given in equation
(\ref{eqn:jeanslengthdef}), by a factor $\pi(4/3)^{1/2}=3.63$. On the other hand, $\leff$ is comparable to the
radius of the fragments we find in our simulations, which have a diameter of
$\lj/2$, corresponding to a radius of $\lj/4$.
Bates' effective Jeans mass, which is the mass within a radius $\leff$, is smaller
than the standard value in equation (\ref{eqn:jeansmassdef}), by a factor $\pi^3/\surd 27\simeq 6.0$.

To estimate the dust temperature at a distance
$\leff$ from a star of luminosity $L$, we follow \citet{Bate2009} and ignore the
frequency dependence of the dust absorption coefficient; $\leff$ is then given by the condition
\beq
L=4\pi\leff^2\ssb T^4,
\eeq
where $\ssb$ is the Stefan-Boltzmann constant. Since $c_s=(kT/\mu\mH)^{1/2}$,
we have
\beq
T=\left(\frac{G\mu\mH\rho L}{3k\ssb}\right)^{1/5}=5.3\left(\frac{\rho}{10^{-19}\mbox{ g cm\eee}}
\cdot\frac{L}{L_\odot}\right)^{1/5}~~~\mbox{K}.
\eeq
\citep{Bate2009} adopted a fiducial luminosity of $150\, L_\odot$, which gives
$T\simeq 15$~K for $\rho\simeq 10^{-19}$~g~cm\eee. This is larger than the
background temperature of 10 K that we have assumed;
the Jeans mass is therefore increased and fragmentation is suppressed.
However, the observed median luminosity in local star-forming regions is far
smaller: \citet{Enoch2009} find a median luminosity of only $1.5\, L_\odot$,
which gives a temperature of only 5.8 K at $\leff$. Since this is less than
the background temperature, we conclude that protostars typically do not suppress fragmentation at densities of order $10^{-19}$~g~cm\eee, which are
characteristic of low-mass star-forming regions. At higher densities, the accretion luminosity can raise the temperature enough to begin to suppress fragmentation in clouds with $T\simeq 10$~K.

\subsection{Fragmentation in Rho Ophiuchus}

Our cloud fragments at about half the Jeans length (i.e. at $\sim$0.1 pc), but then continues to fragment below this point.  Fragmentation at the Jeans length is commonly observed \citep{Blitz1997,Enoch2008}.  In instances where observers have the resolution and sensitivity to resolve fragmentation at scales below the Jeans length, however, even more fragmentation is found \citep{Motte1998, Johnstone2000, Teixeira2007, Chen2010, Bontemps2010}.  Fragmentation can be quantified as a function of scale, $r$.  Given a set of clump locations in a cloud, one can calculate a set of clump pair separations.  Let the differential number of pairs separated by distance $r$ be $d N_{\rm{pair}} = H(r) d \ln{r}$.  The number of clump pair separations for randomly distributed clumps is $H_{\rm{ran}}(r)$.  The two point correlation function, $w$ can be calculated from these quantities \citep{Johnstone2000},
\begin{equation}
w(r) = \frac{H(r)}{H_{\rm{ran}}(r)} - 1.
\end{equation}
The two-point correlation function has been measured for the central parsec of $\rho$ Ophiuchus by \citet{Johnstone2000}. In this measurement, excess fragmentation ($w > 0$) is found below $r \sim 3 \times 10^{4}~\rm{AU}$, similar to the Jeans length of the cloud.  There is a power law fit, $w(r) \propto r^{-0.75}$, in this regime.  \citet{Larson1995} also measured clustering of stars in Taurus and found a power law fit with a break at 8000 AU, but attributed the break to binary stars.  The separation between stars has had time to evolve since the initial fragmentation, so we narrow our focus to comparisons with \citet{Johnstone2000}.

\begin{figure}[!ht]
\center{\includegraphics[width=3.5in]{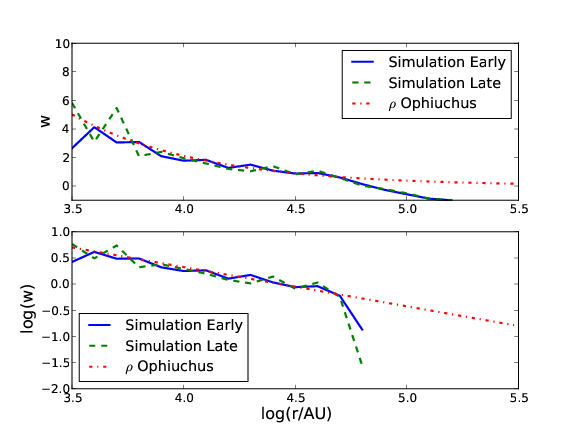}}\caption{Two point correlation function, $w$, for clumps in simulation RW convolved with a 1600 AU Gaussian beam.  The correlation function is calculated at an early time, $t \sim 0.4 t_{\rm{ff}}$ and a late time, $t \sim 0.75 t_{\rm{ff}}$.  The correlation is similar at early and late times, except for the smallest scales, where fragmentation increases over time.  For comparison, the fit to $r < 3 \times 10^{4}$ AU measurements from $\rho$ Ophiuchus is also included.} \label{fig:clump2point}
\end{figure}

To compare our simulations to the observed $w$, we first created optically thin column density maps of our simulations and convolved them with a Gaussian with a FWHM of 1600 AU.  The resolution was chosen to be similar to that from \citet{Johnstone2000}.  We used the Clumpfind algorithm from \citet{Williams1994} on the convolved column density map from simulation RW to obtain a list of clumps and their positions.  To investigate the possibility of time evolution of $w$, this is performed at an early time in the simulation and then again at a late time, $t \sim 0.4 t_{\rm{ff}}$ and $t \sim 0.75 t_{\rm{ff}}$, respectively.  The results are shown in Figure \ref{fig:clump2point}.
As expected, the correlation function drops off above $r \sim 4 \times 10^{4}$ AU, about 2/3 the Jeans length at $\bar{\rho}$ for our simulation.  More remarkably, the correlation function in our simulation matches that measured in $\rho$ Ophiuchus quite well at all scales below this drop off.  The early and late time simulations also generally agree with each other, suggesting there is not much time evolution in $w$.  There is a discrepancy between the two times at $r \sim 5 \times 10^3$ AU.  At these small scales, fragmentation does increase in time, as high density regions have more time to form and fragment.

It should be noted that at $t \sim 0.4 t_{\rm{ff}}$, our stars have not provided very much feedback, and the simulation can be considered solely gravito-hydrodynamic.  Simulations with just hydrodynamics and gravity are scale-free.  The exact match with $\rho$ Ophiuchus is partially due to our choice of cloud parameters to mimic $\rho$ Ophiuchus.  The scale-free results are the break in $w(r)$ at the Jeans length the $w(r) \propto r^{-0.75}$ functional form.

\subsection{Observed Core Mass Functions}
There is a wealth of observations cataloguing masses of cores in star-forming regions \citep{Motte1998, Testi1998, Johnstone2000, Johnstone2001, Motte2001, Beuther2004, Stanke2006, Alves2007, Enoch2008, Sadavoy2010}.  Most of these observations are unable to resolve the small scale fragmentation seen in $\rho$ Ophiuchus, but still provide valuable information.  To recreate the observations, we produced optically thin column density maps of our simulations in all three directions and convolved them with Gaussian beams chosen to match the observations.  We then applied Clumpfind to the convolved column density, similar to our comparison to $\rho$ Ophiuchus.

\begin{figure}[!ht]
\center{\includegraphics[width=3.5in]{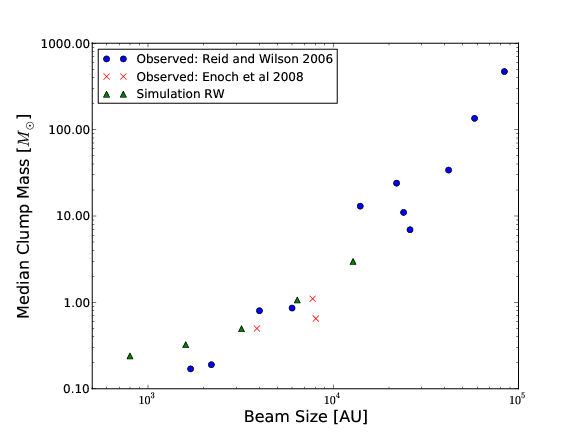}}\caption{Median mass of the CMF found in a cloud as a function of resolution of the observation.  The CMFs from synthetic observations of RW are green triangles.  For comparison, CMFs tabulated in \citet{Reid2006} are included as well as 3 CMFs from \citet{Enoch2008}.  The two lowest points from \citet{Reid2006} are $\rho$ Ophiuchus at different wavelengths.} \label{fig:reidplot}
\end{figure}

The observations have a wide range of beam sizes due to the range in distances to star-forming regions, so we also used a range of beam sizes for comparison.  We find that the CMF derived from our simulated observations is highly sensitive to the beam size used.  As the beam size increases, the smallest cores are no longer detectable and drop out of the CMF.  In addition, tight clusters of cores become unresolved and look like new, much larger cores.  Both effects increase the median clump mass.  The effect of overlapping cores has been explored further in \citet{Kainulainen2009a}.  This sensitivity of the CMF to resolution is seen also in the observations, and shown in Figure \ref{fig:reidplot}.
To gather the median mass of a range of CMFs, we used the data tabulated in \citet{Reid2006}.  All cores from \citet{Reid2006} are detected using Clumpfind.  To supplement that data, we also used the three clouds (Serpens, Perseus and $\rho$ Ophiuchus) from \citet{Enoch2008}.  \citet{Enoch2008} do not rely solely on Clumpfind, but use a method that returns similar results.  Clumpfind has many limitations \citep{Pineda2009, Goodman2009}, but is still useful for the purposes of comparison.  When compared to these observational data, our simulated CMFs match quite well.  The simulated CMF with 3200 AU and 6400 AU beams are interesting in particular, as this is the usual range of telescope beams for nearby clouds.  The extent of our simulation domain (130,000 AU) prevents useful comparisons to more distant observations.  In addition, our base grid resolution, 512 AU, causes discrepancies with observations of cores at small beam sizes.  Bound cores will trigger adaptive refinement and can go to smaller scales, but smaller, ephemeral, cores that would show up in observations are not always captured in the simulations.  This means the simulated median masses at 800 AU and 1600 AU are too high compared to observations which include all cores, bound or not.  This is most apparent when comparing to the observations of $\rho$ Ophiuchus with a beam size of 1600 AU.  The net effect is a flattening in the clump mass versus beam size relation at small beam sizes for the simulated clump observations.

Figure \ref{fig:reidplot} does not include the CMF from the Pipe Nebula \citep{Alves2007}.  The resolution is comparable to the best observations of $\rho$ Ophiuchus, but the median mass is near 1 $M_{\odot}$.  This makes it stand apart from the other observations.  This region has a low density and is not actively forming stars.  In addition, the observed cores are largely unbound \citep{Lada2008}.  The Pipe Nebula should perhaps be considered a pre-star-forming region, unlike Serpens or $\rho$ Ophiuchus.  These pre-star-forming regions have much higher Jeans lengths and masses and have fundamentally different column density distribution functions \citep{Kainulainen2009b}.

\subsection{CMF to IMF relation}

\begin{figure}[!ht]
\center{\includegraphics[width=3.5in]{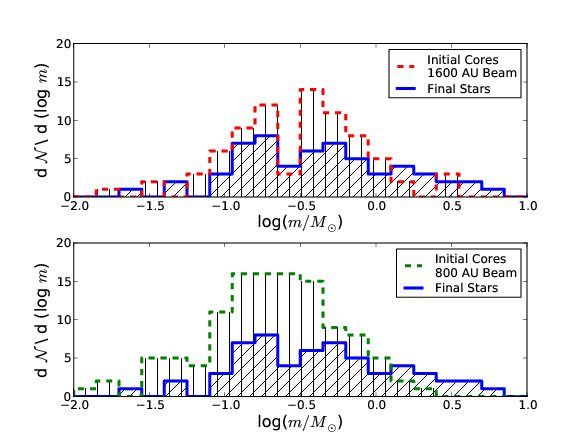}}\caption{Mass functions for both the initial cores found using clumpfind and for the final masses in stars.  The y axis represents total counts and is not normalized.  Top panel:  cores found using a 1600 AU beam size.  Bottom panel:  cores found using an 800 AU beam size.} \label{fig:imfcmf}
\end{figure}

Given that we have identified clumps at the beginning of star formation and have stars at the end of the simulation, a natural question is how the initial clumps relate to the final stars.  We will first consider the cores identified by clumpfind discussed in the previous section.  These cores represent what an observer might find in a similar cloud.  For this comparison, we need to use simulations that have run to completion, so cannot use the radiative simulations.  As a first pass at the correlation between the CMF and IMF, we will start with simulation B, where winds are absent and one might expect the CMF and IMF to overlap. The initial CMF and the final IMF for B are shown in Figure \ref{fig:imfcmf}.  For proper comparison, the mass functions are not normalized, and total counts at each mass are shown.  To avoid triple counting cores in the CMF, the synthetic observations are taken in only one line-of-sight direction instead of all three.  Otherwise these CMFs are the same as those in previous sections.
When focusing on the CMF found with the 1600 AU beam size, the initial CMF and final IMF are well correlated.  The typical mass and total number of objects match well between cores and stars.  Unfortunately, this comparison only holds for the 1600 AU beam size.  The CMF is highly sensitive to beam size, as shown in Figure \ref{fig:reidplot}.  When a slightly smaller, 800 AU, beam size is used, the clumps are too small and too numerous to all correlate with stars.  It should be noted that this analysis is performed using a single clump identification method (Clumpfind) on a crowded field.  It is possible that different methods of clump identification or clouds with well-separated clumps do not have as much sensitivity to telescope resolution in the resulting CMF.  Even without resolution effects, we are considering all cores, bound and unbound.  Low-mass cores are less likely to be bound or to form stars \citep{PhilMyers2009, Padoan2011}, which will skew the CMF to lower masses.  CMFs that only include bound cores should be more correlated with the IMF.

We do not need to limit ourselves to synthetic observations of our simulations and can identify cores from the full 6 dimensional phase space.  This is done using the `find\_clumps' routine in the yt analysis toolkit \citep{Turk2011}.  The algorithm is described in more detail in \citet{Smith2009}.  It uses density contours to return a hierarchy of clumps, where each clump can contain smaller child clumps.  Our simulations have hundreds of local maxima in density large enough to be considered clumps.  At $t = 0.3 t_{\rm{ff}}$, only 40 of these are bound, defined as $|\rm{potential~energy}| > (\rm{thermal~energy} ~+~\rm{kinetic~energy})$.  We will only consider the bound clumps in this analysis.

\begin{figure}[!ht]
\center{\includegraphics[width=3.5in]{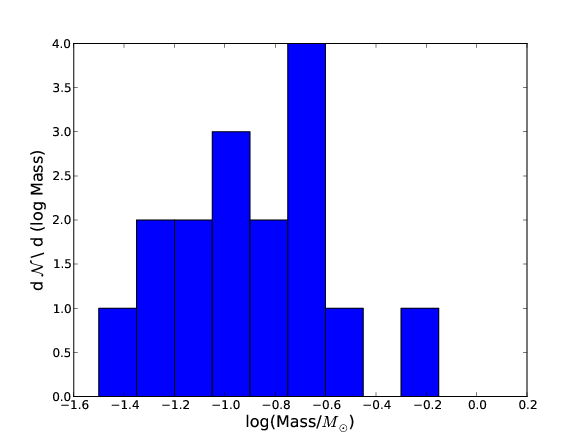}}\caption{Histogram of masses of bound childless sub-cores at $t = 0.4 t_{\rm{ff}}$ for simulation RW.  } \label{fig:boundcmf}
\end{figure}

The bound clumps cover a large range of sizes with the largest clump filling almost the entire box (181 $M_{\odot}$).  This clump contains five child clumps.  Each of these clumps contain many generations of children and grandchildren.  When comparing to the IMF, one should ignore clumps with multiple child clumps, and count the children instead.  Throwing out clumps with masses less than $0.05 M_{\odot}$, there are only 5 bound childless clumps at $t = 0.3 t_{\rm{ff}}$, for a total mass of 2.5 $M_{\odot}$.  At $t = 0.4 t_{\rm{ff}}$, some of the parent clumps split into more children, resulting in 16 bound childless clumps, though the total mass is largely unchanged, at 2.6 $M_{\odot}$.  These clumps are notably smaller than the primary turbulent cores described in \S 3.1 and will be called `sub-cores'.  These sub-cores lead to the burst of star formation at $t = 0.5 t_{\rm{ff}}$ and their mass function is shown in Figure \ref{fig:boundcmf}.
The median mass of these sub-cores is 0.13 $M_{\odot}$.  Even if each sub-core forms exactly 1 star, they do not explain the stellar mass function of the simulation, which eventually has a median mass of 0.3 $M_{\odot}$ when winds and radiation are not included.  This discrepancy can be explained by the $20,000$ AU turbulent cores.  These objects are bound, but they cannot be detected with density contours due to their supersonic turbulence.  Each core has multiple pockets of high density gas.  When using density contours, one sees many smaller unrelated, unbound clumps instead of a few larger bound clumps that correspond to the physical cores.  The cores are visible by eye and should be indentifiable with a more advanced density search.  We identify them by stellar clustering.  The central stellar systems (usually binary systems) in these cores are all much more massive than the sub-cores.  If these central stellar systems are ignored, the median mass of stars moves from 0.3 $M_{\odot}$ to 0.16 $M_{\odot}$, much closer to the median sub-core mass.  The median mass of non-central stars for the case with winds is 0.07 $M_{\odot}$.  The turbulent core and central star properties are summarized in table \ref{tab:cores}.
\begin{table}[scale=0.8]
\begin{flushleft}
\caption{Turbulent Core Properties for Barotropic Simulations.}\label{tab:cores}
\begin{tabular}{|c|c|c|c|c|c|c|}

 \hline
    & \multicolumn{3}{|c|}{Without Winds} & \multicolumn{3}{|c|}{With Winds} \\
  \hline
  $M_{\rm{gas}}$ Initial & $M_{\rm{gas}}$ Final & $M_{*,\rm{total}}$ & $M_{*,\rm{central}}$ &
  $M_{\rm{gas}}$ Final & $M_{*,\rm{total}}$ & $M_{*,\rm{central}}$ \\

  \hline
  3.4 & 3.6 & 5.6 & 4.7 & 2.3 & 1.9 & 1.3\\
  7.0 & 3.1$^a$ & 16.8$^a$ & 9.4 & 2.0$^a$ & 9.6$^a$ & 5.9\\
  3.8 & 2.5 & 5.0 & 3.4 & 1.2 & 1.5 & 1.4\\
  5.0 & 2.7 & 5.1 & 3.9 & 1.6 & 2.9 & 2.2\\
  2.9 & -$^a$ & -$^a$ & 1.0 & -$^a$ & -$^a$ & 0.04 \\
  3.3 & -$^a$ & -$^a$ & 2.7 & -$^a$ & -$^a$ & 0.3 \\

  \hline

\end{tabular}
\end{flushleft}
$^a$ \footnotesize{The last two cores eventually merge with the largest core, making it impossible to measure the final gas and total stellar properties.  The final properties of the largest core are necessarily a sum over the last two cores in addition to the largest core.}
\end{table}

The cores are bound at late times even when only gravity due to the stars is considered, with the kinetic energy in stars approximately half the potential energy of stars in each core.  In addition, the core-to-core velocity dispersion is typically 0.4-0.5 km s$^{-1}$.  This is notably lower than the cloud velocity dispersion, which starts at 1.2 km s$^{-1}$.  There is a strong anti-correlation between velocity and density, as the densest gas occurs at stagnation points in a turbulent flow.  This means that the core-to-core velocity dispersion will naturally be much lower than that of the gas \citep{Offner2009b}.

Once the cores have formed, each core is carved out by the outflows of its own protostellar system.  This yields the core to star efficiency factor, $0.2 < \epsilon_{\rm{core}} < 1.0$.  The amount of mass lost from a spheroidal core can be calculated from the total momentum output and opening angle of the winds \citep{Matzner2000}, but the cores in our simulations are more complicated.

The best way to calculate $\epsilon_{\rm{core}}$ is to compare the mass of stars in simulations with and without winds.  In the simulations with the barotropic equation of state (B and BW), the total, the mean, and the median mass of stars are all approximately 3:1 comparing the non-wind simulation to the wind simulation at any point in time.  This means $\epsilon_{\rm{core}} \simeq 1/3$.  The 2/3 of the core mass that is lost in the presence of winds is mostly core gas that is entrained in the winds.  This was not precisely quantified, and it is possible other mechanisms account for some mass loss, such as unbinding outer core gas when the mass from the interior core is lost.  In the radiative simulations (R and RW), the total mass of stars is also 3:1 comparing the two simulations.  The mean and median masses are 3:1 for simulation RW, but closer to 10:1 in simulation R due to unphysically strong radiative suppression of fragmentation.

In the event that cores are accreting mass, $\epsilon_{\rm{core}}$ is a function of time, and may be larger than 1 if the instantaneous core mass is less than the final stellar system mass \citep{Padoan2011}.  Given that the sub-cores in our simulations are massive enough at early times to create their final stellar systems, the sub-cores probably do not accrete much mass.  It is extremely difficult to track these cores over time, however, so we cannot quantify the amount of sub-core accretion.

\subsection{Turbulent Fragmentation and Competitive Accretion}

It is useful to place the stellar accretion in our simulations in the context of existing star formation models.  Two current popular models are turbulent fragmentation and competitive accretion.  In the turbulent fragmentation model, \citep{Padoan1995, Padoan2002, McKee2003, Padoan2007, Hennebelle2008, Hennebelle2009}, supersonic turbulence in molecular clouds creates many cores.  Each bound core then collapses into a single stellar system (if the structure of the core is dominated by internal turbulence, the resulting model for the formation of the star was termed the ``turbulent core model" by \citealt{McKee2003}).  In this scenario, the mass from each star is accreted almost entirely from its natal core.  In the competitive accretion model \citep{Zinnecker1982, Bonnell1997, Bonnell2001, Bate2005, Bonnell2006}, the bound cores begin with a mass $\sim 0.1~M_{\odot}$.  The molecular cloud undergoes a global collapse and all stars accrete from the entire cloud.  Protostars exhaust their cores at low-masses and then grow by Bondi-Hoyle accretion.  The protostars compete with each other for mass from the host cloud and the dynamics of this competition lead to the IMF.  Roughly speaking, the virial parameter of the cloud decides which model is applicable \citep{Krumholz2005b, Bonnell2006b, Offner2008}.  In clouds with sufficiently sub-virial turbulence, global collapse is possible and competitive accretion prevails.  In virialized clouds, core accretion dominates.  Most simulations of star formation start with virialized clouds, but turbulence quickly dissipates and simulations that do not regenerate turbulence become sub-virial and demonstrate competitive accretion.  Turbulence can be regenerated by protostellar outflows, HII regions, or a cascade from larger scales; in simulations, the cascade can be regenerated by large scale driving.

Our simulations largely agree with turbulent fragmentation models, while introducing a hierarchical aspect of sub-cores within turbulent cores.  We form turbulent cores on the cloud Jeans length and each of those cores forms a central binary or single star with mass roughly equal to the core mass.  Even the smaller stars are formed in their own sub-cores and do not accrete from the cloud at large.  Our turbulent cores do accrete from the larger cloud (increasing their initial mass by $\sim 75\%$ over the course of the simulations), which was not originally part of the core accretion theory, but recent turbulence simulations suggest turbulent cores do accrete from their host cloud \citep{FalcetaGoncalves2011}.

The accretion from the larger cloud onto cores in our simulation is possibly caused by the fact that we have not included turbulent driving from large scales.  Turbulence is regenerated to some extent by protostellar outflows, but this is relatively ineffective in denser gas.  Our simulations are then similar to the undriven simulations in \citet{Offner2008}, which also show accretion onto cores.  The simulations in \citet{Offner2008} with external driving produce cores that do not accrete much from the cloud.  If magnetic fields were included in our simulations, the outflows would couple to much more of the gas \citep{Wang2010}, which would reduce accretion onto the cores.

\subsection{The Role of Stellar Mergers}

The stellar mass functions in our simulations are influenced by the details of our sink particle merger process.  Mergers are necessary because all codes will introduce numerical fragmentation once they can no longer resolve the Jeans length on the finest scale \citep{Truelove1997}.  More stringent sink particle conditions can reduce the number of unwanted sink particles \citep{Federrath2010}, but numerical fragments are unavoidable.  If sink particles are not allowed to merge, these numerical fragments will steal mass from physical fragments and masquerade as real stars, artificially lowering the IMF.  On the other hand, allowing sink particles to merge has a similar effect on the IMF as suppressing gas fragmentation.  If all sink particles that pass near each other are merged together, the particles will consolidate over time.  Eventually, all IMFs look similar to the top heavy IMF from R, where radiation suppressed most fragmentation.  Our decision to only merge protostars with masses less than $0.05 M_{\odot}$, the mass of the first core of a 1 $M_{\odot}$ star, is a compromise between the two extremes of no mergers and all mergers.

This suppression of mergers is seen comparing the IMFs of RW and BW.  The radiative simulation should suppress some fragmentation while the barotropic simulation fragments all the way down to the Jeans mass at the self-opacity limit, $m \sim 0.004 M_{\odot}$.  Nevertheless, they have the same IMF.  The barotropic simulation does fragment much more than the radiative simulation near the resolution limit.  This is seen in the total number of sink particles created, where the barotropic simulation creates 7 times more particles than the radiative simulation.  These particles are nearly all very small in mass and immediately merged.  The net effect of the extra mergers is to suppress fragmentation by combining fragments below the merger radius (128 AU).  The two simulations are nearly identical above this scale and therefore produce the same IMF.

Our choice of merger mass ($m_{\rm{merge}} = 0.05~M_{\odot}$) is based on calculations of first collapse of a solar mass star \citep{Masunaga1998, Masunaga2000}, but the correct mass is not certain.  In addition, SPH simulations with resolutions of a few AU (compared to our 32 AU) move their numerical fragmentation to much smaller scales and do not show stellar mergers of this type \citep{Stamatellos2009, Bate2011}. To investigate the effect of our mass choice, we repeated RW with a merger mass of 0.01 $M_{\odot}$ out to $t = 0.55~t_{\rm{ff}}$.  At this point in the simulations, the total mass in stars is $2~M_{\odot}$.  In the simulation with $m_{\rm{merge}} = 0.05~M_{\odot}$, there are 17 total sink particles; whereas in the simulation with $m_{\rm{merge}} = 0.01~M_{\odot}$, there are 30 total sink particles, even though the total mass in particles is the same.  Given that the final IMF for $m_{\rm{merge}} = 0.05~M_{\odot}$ is at slightly lower masses than the observed IMF, nearly doubling the number of sink particles with $m_{\rm{merge}} = 0.01~M_{\odot}$ would move the IMF to masses less than half the observed value.

\subsection{Scaling}
For most isothermal simulations, there is a free parameter which allows rescaling of the mass, $M$, density, $\rho$, or temperature $T$, while maintaining all of the same dimensionless parameters $\alpha$, $\mathcal{M}$, and $M_J / M$.
When the merger mass is included, $m_{\rm{merge}}$ also scales with $M$.  Now if we scale the simulation to a higher mass, we are also increasing $m_{\rm{merge}}$.  There is some uncertainty in $m_{\rm{merge}}$, but it would be difficult to justify increasing it much more than our current level.  Even increasing $m_{\rm{merge}}$ by a factor of 2 would bring it to uncomfortably close to the characteristic IMF mass of $0.2~M_{\odot}$.  This means the mass scales of our simulation are relatively stationary.  When protostellar winds are included, they introduce a new fixed dimensionless number, the Mach number of the winds $\mathcal{M}_{\rm{wind}}$.  Since the speed of the winds is proportional to the escape velocity from the stellar surface (i.e., $\propto M^{1/2}$ \citealt{McKee2007}), $\mathcal{M}_{\rm{wind}}$ sets the quantity $M/T$.  In practice, this is not a very tight constraint because the wind speed itself is quite uncertain \citep{Downes2007}.
When radiation is important, the luminosity of each star is set by complicated stellar models that depend on $M$ as well as the accretion history.  The time scale, which goes into the accretion rate, is set by $t_{\rm{ff}}$ and therefore $\rho$.  In addition, the resulting radiation-hydrodynamics depends on the temperature.  This uniquely sets all scales in the simulations.  Even when radiation is not dynamically important, we do match the observed protostellar luminosities and cannot change our masses without jeopardizing the agreement.  Using the approximation $L \propto M / t_{\rm{ff}} \propto M^{3/2}$, our cloud mass is constrained to $160~M_{\odot} < M < 200~M_{\odot}$ before it no longer falls in the error bars luminosities.  Cloud masses below the lower limit do not match the observed median luminosity and masses above the upper limit do not match the mean.  Our general match to the IMF, which sets $M$, provides an additional, looser constraint.

\section{Conclusions}

We report the results of several simulations of the formation of a low-mass star-forming cluster, comparable to the central parsec of $\rho$ Ophiuchus.  Our simulations achieve 32 AU resolution using adaptive mesh refinement.  We also include radiation-hydrodynamics and protostellar feedback.  The protostellar feedback represents both protostellar radiation and bipolar outflows.  To isolate the individual effects of radiation and outflow feedback, we perform a suite of 4 simulations: a base simulation with no feedback (i.e., with a barotropic equation of state and no outflows), a radiative simulation with no outflows, an outflow simulation with no radiation and a simulation with both outflows and radiation.  This is the first simulation of a low-mass star-forming cluster with both radiation and protostellar outflows.

When outflows are included in the simulation, they are able to replenish the kinetic energy lost from decaying turbulence.  This new outflow-driven turbulence is fundamentally different than isotropic turbulence driven on large scales, however, as seen in \citep{Nakamura2007}.  This can be measured in the solenoidal to compressional energy ratio, which climbs from 2 to 7 over the course of our simulations.  Hydrodynamic outflow-driven turbulence does not couple well to the high density gas and cannot prevent cores from collapsing.

Both simulations with outflows reproduce the expected mass functions.  The radiative simulation does not finish accreting, but it matches the turbulent core protostellar mass function (PMF) from \citet{McKee2010}.  The barotropic simulation has mostly finished accretion and matches the \citet{Chabrier2005} IMF.  Simulations without winds produce mass functions that are too massive by factors of 3 and 10 for the barotropic and radiative simulations, respectively.

When we compare final stellar masses with and without outflows, we find the outflows remove 2/3 of the mass that would go into stars.  This creates a core efficiency parameter $\epsilon_{\rm{core}} \simeq 1/3$, similar to predictions from \citet{Matzner2000}.  The importance of outflows in our IMF calls into question simulations that produce the observed IMF without outflows (e.g. \citealt{Bate2009, Price2009}).  The outflows do not significantly  affect the overall cloud dynamics, as they have small opening angles and do not couple well to the dense gas in the cores.  It is likely magnetic fields would change that conclusion \citep{Wang2010}.

Outflows also significantly reduce protostellar luminosities.  They reduce the accretion rate onto a protostar by a factor of 3 by disrupting its parent core, and therefore reduce the final mass of the protostar by a factor of 3.  The typical radius of a protostar does not change much in this mass regime, so accretion luminosity is reduced by an order of magnitude.  This luminosity lost this way is much larger than the mechanical luminosity of the outflows themselves.

The reduced luminosities in the simulation with radiation and outflows allows it to match the observed protostellar luminosities of nearby star-forming regions.  This includes the mean and median luminosities as well as the standard deviation of the log luminosity.  The protostellar luminosity function also depends on the time and mass evolution of the protostellar accretion rate\citep{Offner2011}.  We find the accretion rate scales with mass as predicted from core accretion models that include both thermal and nonthermal motions \citep{PhilMyers1992, McKee2003} as opposed to competitive accretion or isothermal sphere models.  We also find the accretion rate must be tapered and is consistent with the linear tapering in \citet{McKee2010}.  Since our models do not resolve disk physics, they do not include episodic accretion. The fact that we nonetheless are able to reproduce the observed protostellar luminosities in regions of low-mass star formation suggests that episodic accretion is not the dominant factor in resolving the luminosity problem for low-mass protostars; this is consistent with the discussion of \citet{Offner2011}, but not with that of \citet{Dunham2010}.

The simulation with radiation and without outflows confirms the finding of \citet{Offner2009} and \citet{Krumholz2011} that protostars can heat their host cloud enough to suppress fragmentation.  When outflows are included, however, the total luminosity of stars drops by an order of magnitude, and radiation is far less effective at suppressing fragmentation.  The simulation with radiation and winds has over twice as many stars as the simulation with radiation but without winds.  When fragmentation is additionally suppressed by merging pre-second-collapse stars at 128 AU, radiation has almost no effect on the resulting mass function.  Thus, radiation may be necessary to suppress fragmentation below $\sim$128 AU for the conditions simulated here, but it does not significantly effect the gas dynamics above those scales.

We are able to recreate the clustering properties of the cores found in $\rho$ Ophiuchus, measured by the two point correlation function, $w(r)$ of observed clumps.  Our simulated observations recreate the overall normalization, the $w(r) \propto r^{-3/4}$ slope below the Jeans length, and the break in the power law at the Jeans length observed by \citet{Johnstone2000}.  This implies our simulation of fragmentation is accurate down to at least 2,000 AU, set by the resolution of the observations.

To investigate the conversion of the observed core mass function (CMF) to the stellar IMF, we create simulated dust maps and find cores using Clumpfind.  The mean core mass in observations systematically increases with telescope resolution, and we are able to recreate those core masses over an order of magnitude of resolutions.  At resolutions typical of observations of nearby star-forming regions, 1600 AU, the early time CMF and the final IMF overlap when outflows are not included.  Outflows shift the IMF to lower masses while retaining the original shape.  When we simulate observations with better resolutions, however, the CMF changes and no longer matches the IMF.  This suggests that current observations suggesting a CMF to IMF correspondence could change at higher resolutions.  It should be noted that this analysis was performed with Clumpfind on a simulated cloud with closely spaced cores.  It is possible that different methods of clump identification or a cloud with more separation between cores would produce a CMF that is not as sensitive to telescope resolution.

In sum, we have used ORION, an adaptive mesh refinement (AMR) gravito-radiation-hydrodynamics code, to simulate low-mass star formation in a turbulent molecular cloud in the presence of protostellar feedback.  Our results for the most realistic simulation, which includes both radiation and feedback, are consistent with observations of the protostellar luminosity function, the core mass function, the two-point correlation function of cores in $\rho$ Ophiuchus, and the protostellar and initial stellar mass functions.  We find that protostellar radiation does not affect fragmentation below 128 AU and that protostellar outflows do not significantly affect large-scale cloud dynamics, serving only to reduce the mass in cores for a core to star efficiency, $\epsilon_{\rm{core}} \simeq 1/3$.  Lastly, we find the accretion histories of our stars match core accretion models that include both thermal and turbulent motions and appear to be inconsistent with competitive accretion or isothermal sphere accretion.  The inclusion of magnetic fields would increase the coupling between outflows and dense gas, possibly changing the large-scale cloud dynamics.  With an increase in resolution, it would be possible to explicitly model protostellar mergers prior to second collapse, as well as investigate the small-scale regime where protostellar radiation suppresses fragmentation.

\begin{acknowledgments}
The authors acknowledge helpful discussions with Stella Offner, Mark Krumholz and Andrew Cunningham and improvements suggested by our anonymous referee and Philip Myers. This research has been supported by the NSF through grants AST-0908553 (CEH, CFM and RIK) and CNS-0959382 (RTF), a Spitzer Space Telescope Theoretical Research Program grant (CFM), and AFOSR DURIP Grant FA9550-10-1-0354 (RTF).  Support for this work was also provided by the US Dept. of Energy at LLNL under contract DE-AC52-07NA (RIK).  Support for computer simulations was provided by an LRAC grant from the NSF through Teragrid resources and NASA through grants from the ATFP.  This research used resources of the National Energy Research Scientific Computing Center, which is supported by the Office of Science of the U.S. Department of Energy under Contract No. DE-AC02-05CH11231.
\end{acknowledgments}

\end{document}